\documentclass[aps,prd,onecolumn,groupedaddress,preprintnumbers,superscriptaddress,noshowpacs,nofootinbib]{revtex4}
\usepackage{amsmath,amssymb,latexsym}
\usepackage[dvipdfmx]{graphicx}
\usepackage{color}

\begin{document}

\title{
Energy extraction from Kerr black holes by rigidly rotating strings
}

%


\author{Shunichiro Kinoshita}
\email{kinoshita@phys.chuo-u.ac.jp}
\affiliation{
Department of Physics, Chuo University, 
Kasuga, Bunkyo-ku, Tokyo 112-8551, Japan
}
\author{Takahisa Igata}
\email{igata@rikkyo.ac.jp}
\affiliation{Department of Physics, Rikkyo University, Toshima, Tokyo 175-8501, Japan
}
\author{Kentaro Tanabe}
\email{ktanabe@rikkyo.ac.jp}
\affiliation{Department of Physics, Rikkyo University, Toshima, Tokyo 175-8501, Japan
}

\preprint{RUP-16-27}

\date{\today}

\begin{abstract}
 In this paper, we show that 
 a rigidly rotating string can extract the rotational 
 energy from a rotating black hole.
 We consider Nambu-Goto strings stationary with respect to
 a corotating Killing vector with an uniform angular velocity
 $\omega$ 
 in the Kerr spacetime.
 We show that a necessary condition of the energy-extraction process is
 that an effective horizon on the string world sheet, which corresponds
 to the inner light surface, is inside the ergosphere of the Kerr
 black hole and the angular velocity $\omega$ is less than that of the
 black hole $\Omega_\mathrm{h}$. 
 Furthermore, 
 we discuss global configurations of such strings in both of a slow-rotation
 limit and the extremal Kerr case.
\end{abstract}

\maketitle

 \section{Introduction}

 It is well known that the rotational energy of black holes can be
 extracted, and 
 various mechanisms of the energy extraction have been proposed: 
 the Penrose process caused by
 a particle fragmentation~\cite{Penrose:1969pc,Penrose:1971uk},
 superradiance with amplifying incident waves for various fields~\cite{zel1,Bardeen:1972fi,staro1,staro2,Teukolsky:1974yv},
 the Blandford-Znajek process in electromagnetic fields around a rotating
 black hole~\cite{Blandford:1977ds}, and so on.%
 \footnote{ For example, the magnetic Penrose
 process~\cite{Wagh:1986tsa}. See also Ref.~\cite{Brito:2015oca}
 and references therein. }
 Extracting energy plays an
 important role in astrophysics as well as in general relativity.

 In this paper, 
 we show that
the rotational energy can be extracted from a rotating black hole by
 a rigidly rotating string twining around it.
 Strings are interesting objects in several contexts; for example, 
 in cosmology cosmic strings will be important probes of the early universe, 
 and in string theory, 
 of course, strings themselves are
 so fundamental to construct the theories. 
 The current system may be fascinating 
 as a simple example of interacting systems of 
 a string and a black hole 
 and furthermore 
 various extended objects and black objects~\cite{Gubser:2006bz,Frolov:2006tc,Mateos:2006nu}. 
 In addition, 
 this is expected to be a toy model of magnetic flux around 
 a rotating black hole. 
 As is clear from the stress-energy tensor of the Maxwell field,
 magnetic flux has magnetic tension.
 To study some phenomena in which magnetic tension is essential but magnetic
 pressure is not so significant, 
 one may substitute magnetic flux with strings.
 Energy-extraction mechanism by strings from this perspective were
 discussed in Refs.~\cite{Semenov:2000vb,Semenov:2001ev}.
 
 We consider stationary Nambu-Goto strings with respect to
 a corotating Killing vector characterizing a frame rotating
 with a constant angular velocity $\omega$,
 which is a linear combination of stationary and axisymmetric Killing
 vectors, 
 in the Kerr spacetime.
 Such rigidly rotating 
 strings in stationary axisymmetric spacetimes
 were studied in Ref.~\cite{Frolov:1995vp,Frolov:1996xw}, 
 where for $\omega=0$, particularly, explicit analytic solutions were
 obtained.
 Here, we will focus on the induced geometry on the string world sheet in
 order to reveal the nature of the energy extraction in detail.
 
 In general, if a system is assumed to be rigidly rotating, 
 the locus where its rotational velocity will 
 exceed the speed of light 
 emerges.
 The surface on which the velocity coincides with the  speed of light is called
 as ``light cylinder'' or ``light surface'' more generally. 
 In stationary cases, 
  it becomes the stationary limit surface at which the norm of the 
corotating 
 Killing vector is zero and beyond which the
 Killing vector becomes spacelike.

 Similarly, for a rigidly rotating 
 string such surface will be reflected
 in an effective Killing horizon on the induce geometry of the string
 world sheet.
 Because dynamics of the string is governed by its induced metric, the
 effective horizon determined by the induced metric acts as a causal
 boundary for the stationary region on the world sheet such that the
 stationary Killing vector is timelike with respect to the induced metric.
 However, whether an effective horizon exists or not depends on the actual
 configuration of the string determined by the equations of motion.
 If the string configuration does not reach the light surface, for
 instance, no effective horizon emerges on the string world sheet.
 The aim of this paper is to explore a rigidly rotating 
 string on which the effective horizon exists with an energy flow.
 This implies that the string should be regularly passing through the
 light surface and extending far beyond.
 It turns out that such strings are characterized by two parameters:
 the angular velocity and the angular momentum flux 
 (i.e., torque).

 The reason why the effective horizon is significant is as follows.
 Even if the string is extending far beyond, 
 a segment of the string beyond
 the effective horizon can never affect causally the other segment
 remaining in the stationary region. 
 This means that we should be unconcerned about the segment of the string
 beyond the effective horizon to solve string dynamics in the stationary region.
 The string is regarded as 
 an 
 ``open'' string, which has different end points, in
 its causal patch where dynamics can be determined by initial conditions
 prepared in the stationary region.
 If there is energy flux on this ``open'' string, the energy can be
 transferred 
 from the inner
 end point near the rotating black hole to the outer end point (or
 infinity) far from the black hole.
 Thus, we accomplish net energy transfer across the stationary region
 causally connected on the string world sheet.

 The paper is organized as follows.
 In Sec.~\ref{sec:rigidly_rotating_string} we examine the induced
 geometry of rigidly rotating 
 strings and 
 the conditions that there exists
 an effective horizon.
 As a result, we obtain a parameter space of the angular velocity and
 the angular momentum flux in which physically reasonable string
 configurations can exist and show that energy-extraction process 
 occurs.
 We discuss global structures of such strings in Sec.~\ref{sec:global}.
 In a slow-rotation approximation, where both of a black hole and
 a string are slowly rotating, 
 we analytically study global solutions, and
 then in the extremal Kerr case we show various global configurations obtained by
 numerically solving the equations of motion.
 Unless otherwise specified, Newton's constant $G_\mathrm{N}$ and
 the speed of light $c$ are set to unity in this paper.

 
 \section{rigidly rotating string in the Kerr spacetime}
 \label{sec:rigidly_rotating_string}
 
 \subsection{Effective horizon and regularity condition}
 \label{sec:2A}

 The metric of 
 the Kerr spacetime in the Boyer-Lindquist coordinates 
  is given by 
 \begin{equation}
  \begin{aligned}
   g_{\mu\nu}dx^\mu dx^\nu
   = - dt^2 + \frac{2Mr}{\Sigma} (dt - a \sin^2\theta d\phi)^2
   + \frac{\Sigma}{\Delta} dr^2
   + \Sigma \,d\theta^2 + (r^2 + a^2)\sin^2\theta \,d\phi^2 ,
  \end{aligned}
 \end{equation}
 where 
 \begin{equation}
  \Sigma = r^2 + a^2 \cos^2 \theta , \quad
   \Delta = r^2 + a^2 - 2Mr .
 \end{equation}
 The larger root of $\Delta(r) = 0$ yields the radius of the event horizon 
 $r_\mathrm{h}\equiv M+\sqrt{M^2 - a^2}$.
 The angular velocity of the event horizon is 
 $\Omega_\mathrm{h} \equiv a/(a^2 + r_\mathrm{h}^2)$.
 The radius of the 
 ergosphere, characterized by $g_{tt}=0$, is  
 $r_\mathrm{ergo}(\theta) \equiv M + \sqrt{M^2 - a^2 \cos^2\theta}$.
 Note that the greek indices, $\mu$ and $\nu$,
 are used for spacetime components.

 In this background, we consider the stationary rotating string of which
 the world sheet is tangent to the corotating 
 Killing vector 
 \begin{equation}
  \xi^\mu = (\partial_t)^\mu + \omega (\partial_\phi)^\mu, 
 \end{equation}
 where a constant $\omega$ is the angular velocity of the rotating
 frame.
 Thus, the string is rigidly rotating with the angular velocity $\omega$.
 This string is embedded as 
 \begin{equation}
  \phi=\omega t + \varphi(\sigma) , \quad
   r=r(\sigma) , \quad \theta=\theta(\sigma) ,
 \end{equation}
 where we have synchronized the time coordinate of the string world sheet
 with $t$ in the Boyer-Lindquist coordinates.
 The induced metric on the string world sheet is given by 
 \begin{equation}
  \begin{aligned}
   h_{ab} d\sigma^a d\sigma^b 
   =& 
   - \left[1 - \frac{2Mr}{\Sigma}(1 - \omega a\sin^2\theta)^2
   - \omega^2 (r^2+a^2) \sin^2\theta
   \right]dt^2 
   \\
   & + 2 \varphi' \sin^2\theta 
   \left[\omega (r^2+a^2) - \frac{2Mr}{\Sigma}a(1-\omega a\sin^2\theta)
   \right] dt d\sigma \\
   &+ \left[
   \Sigma \left(\frac{r'^2}{\Delta} + \theta'^2\right)
   + \frac{(r^2+a^2)^2 - a^2 \Delta \sin^2\theta}{\Sigma} \sin^2\theta\,
   \varphi'^2
   \right] d\sigma^2,    
  \end{aligned}
 \end{equation}
 where the prime denotes derivative 
 with respect to $\sigma$.
 Note that the latin indices, $a$ and $b$,
 are used for the world sheet components.
 We have defined the induced metric as 
 $h_{ab} \equiv g_{\mu\nu}\partial_a X^\mu\partial_b X^\nu$, where
 $X^\mu(\sigma^a)$
 denote the embedding functions of the string. 
 This metric admits the stationary Killing vector
 $\xi^a = (\partial_t)^a$ that is induced from $\xi^\mu$. 
 Indeed, the Killing vectors $\xi^\mu$ and $\xi^a$ with respect
 to the spacetime metric and the induced metric are directly related to
 each other as 
 $\xi^a \partial_a X^\mu = (\partial_t)^\mu + \omega (\partial_\phi)^\mu
 = \xi^\mu$.
 It turns out that the norm of $\xi^a$ would vanish.
  A locus of such effective (Killing) horizons on the string world sheet
 is determined by 
 \begin{equation}
  F(r,\theta) \equiv - h_{tt} 
   = 1 - \frac{2Mr}{\Sigma}(1 - \omega a\sin^2\theta)^2
   - \omega^2 (r^2+a^2) \sin^2\theta
   = 0.
   \label{eq:horizon_condition}
 \end{equation}
 Its condition can be rewritten as 
 \begin{equation}
  \Delta(r)
   = \frac{[a - (a^2+r^2)\omega]^2}
   {(1 - a\omega \sin^2\theta)^2}\sin^2\theta .
   \label{eq:horizon_condition2}
 \end{equation}
  This condition, indeed, implies that the norm of the Killing
  vector with respect to the spacetime metric vanish, namely
  $g_{\mu\nu}\xi^\mu\xi^\nu = 0$, as well as the induce metric. 
  Therefore, 
  the surfaces that $F(r,\theta)=0$ represents 
  are nothing but
  stationary limit surfaces in terms of 
  the rigid rotation with the angular velocity $\omega$.
  In general, $F(r,\theta)=0$ has two positive roots in terms of $r$
  like a metric of a
  black hole with cosmological constant. 
  The inner surface and the outer one correspond to so-called inner light
  sphere and outer light cylinder, respectively. 
  Note that, if $\omega$ is sufficiently large, the two light surfaces
  will merge into one connected light surface.
  In this case the number of positive roots of $F(r,\theta)=0$ can be
  less than $2$ for a given $\theta$.%
  \footnote{We thank Chulmoon Yoo for pointing out it. }

  We consider the rigidly rotating Nambu-Goto string. 
  The Nambu-Goto action for the string is 
  \begin{equation}
   S = - \int dt d\sigma \sqrt{-h} = \int dt d\sigma \mathcal{L} ,
    \label{eq:NGaction}
  \end{equation}
  where 
  $h$ is the determinant of $h_{ab}$, and 
  we have defined 
  \begin{equation}
   \begin{aligned}
    \mathcal{L} \equiv& - \sqrt{F
    \,\Sigma  
    \left(\frac{r'^2}{\Delta} + \theta'^2\right)
    + \Delta \sin^2\theta\, \varphi'^2} .
   \end{aligned}
   \label{eq:Lagrangian}
  \end{equation}
  For simplicity, we have omitted an overall factor in the above 
  Lagrangian density, because it is irrelevant to dynamics of the
  strings.
  It means that the tension of the strings has been set to unity, 
  or dynamics of the strings with a unit tension has been described. 
  Since this action does not depend on $\varphi(\sigma)$ explicitly, we
  have a conserved quantity given by 
  \begin{equation}
   q \equiv \frac{\partial \mathcal{L}}{\partial \varphi'} = 
    \frac{\Delta \sin^2\theta}{\mathcal{L}} \varphi' .
    \label{eq:def_q}
  \end{equation}
  The sign of $q$ is directly connected with the sign of
  $\varphi'$ because $\Delta \ge 0$ outside the event horizon:
  $q>0$ for $\varphi'<0$ and $q<0$ for $\varphi'>0$.
  This quantity 
  is interpreted as the angular momentum flux
  flowing on the string world sheet
  outwardly,%
  \footnote{Here, ``outward'' 
  means $r'>0$.}
  and it is written as
  \begin{equation}
   \begin{aligned}
    q =& \sqrt{-h}T^{\sigma a}\partial_a X^\mu (\partial_\phi)^\nu
    g_{\mu\nu} \\
    =& - \sqrt{-h}
    \left[h^{\sigma t} g_{t\phi}
    + (h^{\sigma t}\omega + h^{\sigma\sigma}\varphi')g_{\phi\phi}\right] ,
   \end{aligned}
  \end{equation}
  where $T^{ab} = -h^{ab}$ is 
  the energy-momentum tensor on the world sheet for the Nambu-Goto string.
  Similarly, the outward energy flux is written as
  \begin{equation}
   \begin{aligned}
    \omega q =& - \sqrt{-h} T^{\sigma a} \partial_a X^\mu (\partial_t)^\nu
    g_{\mu\nu} \\
    =& 
    \sqrt{-h}\left[
    h^{\sigma t} g_{tt}
    + (h^{\sigma t}\omega + h^{\sigma\sigma}\varphi')g_{\phi t}
    \right].
   \end{aligned}
  \end{equation}
  Obviously, these angular momentum flux and energy flux are respectively
  associated with the Killing vectors in the Kerr spacetime, namely 
  $(\partial_\phi)^\mu$ and $(\partial_t)^\mu$, so that
  they are conserved in terms of $\sigma$.  
  Note that, if one wants to consider a general value $\mu$ of the
  tension rather than unity, one should replace $q$ with $\mu q$. 
 
 Although the locus of the effective 
 horizon is determined by
 $F(r,\theta)=0$,
 whether the effective horizon does actually exist or
 not depends on the configuration of the string given by the equations
 of motion.
 Let us see a condition for the effective horizon to exist regularly on the
 string world sheet.
 We can rewrite Eq.~(\ref{eq:def_q}) as  
 \begin{equation}
  \frac{r'^2}{\Delta} + \theta'^2 =
   \frac{\Delta \sin^2\theta (\Delta \sin^2\theta - q^2)}{q^2\Sigma F} 
   \varphi'^2 .
   \label{eq:energy_eq}
 \end{equation}
 The left-hand side of the above equation cannot 
 be negative 
 outside the event horizon, i.e., $\Delta > 0$.
 On the other hand, the denominator in the right-hand side 
 can change
 its sign beyond 
 the effective horizon $F(r,\theta)=0$.
 Hence, the numerator must be zero at the effective horizon and 
 must change
 its sign for the
 configuration of the string to extend regularly 
 beyond the effective horizon.
 At the effective horizons 
 $F(r,\theta)=0$, we have the regularity
 condition 
 \begin{equation}
  q^2 = \Delta \sin^2\theta.
   \label{eq:regularity_condition}
 \end{equation}
 If the above condition is not satisfied at $F(r,\theta)=0$, then the induced
 geometry becomes 
 singular there and the locus determined by $F(r,\theta)=0$ is a singularity
 rather than a horizon. 
 Note that, 
 if we regard $\varphi$ as a ``time'' coordinate and
 $(dr/d\varphi,d\theta/d\varphi) = (r'/\varphi',\theta'/\varphi')$ as a
 ``velocity,''
 we can
 interpret 
 Eq.~(\ref{eq:energy_eq}) as an equation of mechanical
 energy for a particle
 in an effective potential 
 on two-dimensional curved space.
 In that sense the condition of Eq.~\eqref{eq:regularity_condition} 
 implies
 ``turning point'' of the effective potential.

Now, we are interested in regular solutions of the string
passing through the effective horizon.
Therefore, we will require the string to satisfy the both conditions of 
Eqs.~(\ref{eq:horizon_condition}) and (\ref{eq:regularity_condition}). 
Solving the two conditions outside 
the event horizon
($r>r_\mathrm{h}$), we have two branches of solutions 
in terms of $r$ (see Appendix~\ref{app:derivation}). 
 The locus of the effective horizon 
 is 
 \begin{equation}
  r^\pm_\mathrm{eff}(\omega,q) = \frac{M}{1 \pm q\omega}
   + \sqrt{\left(\frac{M}{1 \pm q\omega}\right)^2 - a(a \mp q)},
   \label{eq:reffpm}
 \end{equation}
 and 
 \begin{equation}
  \sin\theta^\pm_\mathrm{eff} = |q|/[\Delta(r^\pm_\mathrm{eff})]^{1/2} ,
 \end{equation}
 where 
 $q[a-(a^2+r^2)\omega]>0$ is satisfied if the upper sign is chosen
  and $q[a-(a^2+r^2)\omega]<0$ is satisfied if the lower sign is chosen.
 Note that, for simplicity, we have restricted the domain to the interval
 $0\le\theta\le\pi/2$ because the current system 
 is symmetric under 
 $\theta \to \pi - \theta$. 
 Since the branches of $r^\pm_\mathrm{eff}$ are related to each other in
 $q\to -q$ such as
 $r^-_\mathrm{eff}(\omega,q) = r^+_\mathrm{eff}(\omega,-q)$,
 we shall focus on $r^+_\mathrm{eff}(\omega,q)$ in what follows.

 In order to satisfy 
 $0\le\sin^2\theta^+_\mathrm{eff}\le 1$, it turns out
 that $\omega$ lies in the 
 intervals 
 \begin{equation}
 \begin{aligned}
   &\omega_{0}(q)<\omega \le \omega_{\pi/2}(q) \quad (q>0),\\
   &\omega_{\pi/2}(q)\le \omega < \omega_{0}(q) \quad (q<0),
  \end{aligned}
  \label{eq:allowed_region}
 \end{equation}
 where
 \begin{align}
  &\omega_{0}(q) \equiv - \frac{1}{q},
  \label{eq:omega_axis}
  \\
  &\omega_{\pi/2}(q) \equiv
  \frac{a-q}{2M^2 -q(a-q) + 2M \sqrt{M^2 -(a^2-q^2)}} .
  \label{eq:omega_equatorial}
 \end{align} 
 One of the bounds for the angular velocity $\omega_0(q)$ corresponds to 
 $r^+_\mathrm{eff} \to \infty$ and $\theta^+_\mathrm{eff} \to 0$, and
 the other $\omega_{\pi/2}(q)$ corresponds to 
 $r^+_\mathrm{eff} = M+\sqrt{M^2-(a^2-q^2)}$ 
 and
 $\theta^+_\mathrm{eff} = \pi/2$.
 If $q=0$, 
 the bound of the angular velocity of the string, $\omega_{\pi/2}(q)$, 
 and the radius of the
 effective horizon, $r^+_\mathrm{eff}$,
 will coincide with the angular velocity
 $\Omega_\mathrm{h}$ and radius $r_\mathrm{h}$ of the event horizon
 on the equatorial plane, respectively. 
 When $\omega = 0$, the radius of the effective horizon 
 coincides with that of the ergosphere, namely
  $r^+_\mathrm{eff}(0, q)=r_\mathrm{ergo}(\theta^+_\mathrm{eff})$.

 It is worth noting that, because we can find that 
 the sign of $\partial r^+_\mathrm{eff}/\partial\omega$ depends only on 
 the sign of $q$ (see Appendix~\ref{app:derivation}),
 both of $r^+_\mathrm{eff}(\omega,q)$ and
 $\theta^+_\mathrm{eff}(\omega,q)$ are monotonic functions of $\omega$
 for a given $q$.
 The regions determined by Eq.~(\ref{eq:allowed_region}) 
 are whole regions 
 allowed for $(r^+_\mathrm{eff},\theta^+_\mathrm{eff})$ to exist
  (we will later depict the regions in
 Figs.~\ref{fig:parameter_space} and \ref{fig:sketch}).

  \subsection{Equations of motion and boundary conditions}
  
  We have seen the position of the string in the spacetime
  is determined so that the
  configuration of the string should regularly extend over the effective
  horizon on the induced geometry.
  We will show that the regularity conditions can determine the first
  derivative, also.
  It follows that we can obtain boundary conditions enough to solve the
  equations of motion for the string at the effective horizon.
  Once we have required rigidly rotating strings to enter into the
  effective horizon, their boundary conditions are determined.
  
  The action (\ref{eq:NGaction}) still has the degree of freedom arising from 
  coordinate
  transformation of the spatial world sheet coordinate $\sigma$.
  This means that one can choose $\sigma$ 
  for one's own convenience.
   With taking $\sigma$ as an affine parameter
  (i.e., $\mathcal{L} = -1$), the equations of motion for $r(\sigma)$,
  $\theta(\sigma)$, and $\varphi(\sigma)$ are 
  \begin{align}
    &\frac{F\Sigma}{\Delta} r''
    + \frac{1}{2}\left(\frac{F\Sigma}{\Delta}\right)_{,r} r'^2
    + \left(\frac{F\Sigma}{\Delta}\right)_{,\theta} r'\theta'
    - \frac{(F\Sigma)_{,r}}{2} \theta'^2
    - \frac{\Delta_{,r}}{2}\sin^2\theta \varphi'^2 = 0 , 
       \label{eq:EOM1affine}
       \\
    &F\Sigma \theta''
    - \frac{(F\Sigma)_{,\theta}}{2\Delta} r'^2
    + (F\Sigma)_{,r} r'\theta'
    + \frac{(F\Sigma)_{,\theta}}{2} \theta'^2
    - \Delta \sin\theta\cos\theta \varphi'^2 = 0,
   \label{eq:EOM2affine}
  \end{align} 
  and 
 \begin{equation}
  \Delta \sin^2\theta \varphi' = - q ,
   \label{eq:EOM3affine}
 \end{equation}
 which corresponds to Eq.~\eqref{eq:def_q}.
 Note that we have used a comma to denote a partial 
 derivative 
 with
 respect to a spacetime coordinate, such as $F_{,r} = \partial_r F$.

 Since $F(r,\theta)$ in the equations of motion
 vanishes at the effective horizon, we should require conditions for the
 solutions to be regular.
 At $\sigma=\sigma_0$ such that
 $r(\sigma_0)=r_\textrm{eff}^\pm$ and
 $\theta(\sigma_0)=\theta_\mathrm{eff}^\pm$, the regularity for the
 equations of motion yields
 \begin{align}
  \left.\left(
  \frac{F_{,r}}{2} r'^2 + F_{,\theta} r'\theta' - \frac{F_{,r}}{2}\Delta
  \theta'^2
  \right)\right|_{\sigma=\sigma_0}
  =& \left.\frac{\Delta_{,r}}{2\Sigma}\right|_{\sigma=\sigma_0} ,\\
  \left.\left(
  - \frac{F_{,\theta}}{2\Delta} r'^2 + F_{,r} r'\theta' +
  \frac{F_{,\theta}}{2}\theta'^2
  \right)\right|_{\sigma=\sigma_0} 
  =& \left.\frac{\sqrt{\Delta - q^2}}{q \Sigma}\right|_{\sigma=\sigma_0} ,
 \end{align} 
 where we have used
 $\sin\theta_\mathrm{eff}^\pm = |q|/[\Delta(r_\mathrm{eff}^\pm)]^{1/2}$ .
 If 
 $(\Delta_{,r} F_{,r} - 2F_{,\theta} \cot\theta)|_{\sigma=\sigma_0} > 0$,
 we obtain boundary
 conditions for the first 
 derivatives 
 of $r(\sigma)$ and $\theta(\sigma)$ as 
 \begin{align}
  &r'^2|_{\sigma=\sigma_0}
  = \left.\left[
  \frac{\Delta_{,r} F_{,r} - 2F_{,\theta} \cot\theta}
  {2\Sigma(F_{,r}^{2} + F_{,\theta}^2 \Delta^{-1})}
  + \frac{1}{2\Sigma}\sqrt{\frac{\Delta_{,r}^{2} + 4\Delta\cot^2\theta}
  {F_{,r}^{2} + F_{,\theta}^{2}\Delta^{-1}}}
  \right]\right|_{\sigma=\sigma_0} ,\\
  &\theta'^2|_{\sigma=\sigma_0}
  = \left.\left[
  - \frac{\Delta_{,r} F_{,r} - 2F_{,\theta} \cot\theta}
  {2\Delta\Sigma(F_{,r}^2 + F_{,\theta}^2 \Delta^{-1})}
  + \frac{1}{2\Delta\Sigma}\sqrt{\frac{\Delta_{,r}^2 + 4\Delta\cot^2\theta}
  {F_{,r}^2 + F_{,\theta}^2\Delta^{-1}}}
  \right]\right|_{\sigma=\sigma_0} ,
 \end{align} 
 where the relative sign between $r'(\sigma_0)$ and $\theta'(\sigma_0)$
 should be positive
 because of
 $\Sigma(F_{,r}^2+\Delta F_{,\theta}^2)r'\theta'|_{\sigma=\sigma_0} =
 (\Delta_{,r} F_{,\theta} + 2\Delta F_{,r}\cot\theta)|_{\sigma=\sigma_0}>0$.
 Note that we can solve the equations of motion outward or inward from
 the effective horizon $\sigma=\sigma_0$ when we choose
 $r'(\sigma_0)>0$ or $r'(\sigma_0)<0$ respectively.

  \subsection{Parameter regions of physical solutions: energy extraction}

  Provided that parameters $\omega$ and $q$ are 
  given in the allowed regions, we have two sets of the
  boundary conditions $\{r, \theta, r', \theta'\}$ for each branch of 
  $r^\pm_\mathrm{eff}$ and can obtain regular string solutions at least
  near the effective horizon.
  Since the angular momentum flux $q$ and the energy
  flux $\omega q$ have been defined to be positive when their directions are outward,
  $q>0$ and $\omega q>0$ can respectively describe processes of extracting the
  angular momentum and the energy from the black hole.   
  If one is interested in the energy extraction for example, the string solutions
  with $\omega q > 0$ are important.
  However, even though the string configurations are regular, all of
  them cannot 
  describe physical processes.
  Only one of the branches 
  of solutions
  can describe physically reasonable 
  process, while the other describes unphysical but time-reversal process.

  As a simple and intuitive manner to discriminate the physical process,
  we will argue total thermodynamic system constituted of the string and the
  rotating black hole.
  We suppose the total energy and total angular momentum
  conservation are satisfied for the total system.
  Then, we have 
  \begin{equation}
   \frac{dM_\mathrm{BH}}{dt} + \omega q = 0, \quad
    \frac{dJ_\mathrm{BH}}{dt} + q = 0 ,
  \end{equation}
  where $M_\mathrm{BH}$ and $J_\mathrm{BH}$ are the mass and angular
  momentum of the black hole.
  The first law of 
  black hole thermodynamics%
  \footnote{ The first law of black hole 
  thermodynamics for the Kerr black hole is given by
  $dM = T dS + \Omega_\mathrm{h} dJ$, where the 
  temperature $T$, entropy $S$, angular momentum $J$, and 
  angular velocity $\Omega_\mathrm{h}$ are defined by
  \begin{equation*}
   T = \frac{\Delta_{,r}(r_\mathrm{h})}{4\pi(r_\mathrm{h}^2+a^2)} ,\quad
    S = 2\pi M r_\mathrm{h} ,\quad
    J = aM ,\quad
    \Omega_\mathrm{h} = \frac{a}{r_\mathrm{h}^2 + a^2} .
  \end{equation*}
  }
  yields 
  \begin{equation}
   \frac{dS_\mathrm{BH}}{dt} = \frac{1}{T}
    \left(\frac{dM_\mathrm{BH}}{dt}
     - \Omega_\mathrm{h}\frac{dJ_\mathrm{BH}}{dt}\right)
    = \frac{q}{T}(\Omega_\mathrm{h} - \omega) ,
  \end{equation}
  where $T$ is the Hawking temperature.
  Since physical processes should satisfy $dS_\mathrm{BH}/dt \ge 0$,
  this implies that 
  $\omega \le \Omega_\mathrm{h}$ for $q\ge 0$ and
  $\omega \ge \Omega_\mathrm{h}$ for $q \le 0$.
  Therefore, the physically reasonable 
  solutions seem to be those of 
  the $r^+_\mathrm{eff}$ branch.

  The induced geometry on the string world sheet offers further insights
  into criteria for determining physical process. 
  Now, $(t,\sigma)$ component of the induced metric can be rewritten as  
  \begin{equation}
   \begin{aligned}
    h_{t\sigma} 
    =& \frac{\varphi'\sin^2\theta}{\Sigma}
    \left\{
    \Delta a(1-\omega a\sin^2\theta) - (r^2+a^2)[a-\omega(r^2+a^2)]
    \right\} .
   \end{aligned}
  \end{equation} 
  At the effective horizon, $\sigma=\sigma_0$, we evaluate it as 
  \begin{equation}
   h_{t\sigma}|_{\sigma=\sigma_0}
    = -
    \left.
     \mathcal{L}
     \frac{q[a-(a^2+r^2)\omega]}
     {\Delta(r)(1-\omega a\sin^2\theta)}
    \right|_{\sigma=\sigma_0}
    .
  \end{equation} 
  Since $\mathcal{L}<0$, $\Delta(r_\mathrm{eff}^\pm) > 0$, and $1-\omega a\sin^2\theta_\mathrm{eff}^\pm >0$ (see Appendix~\ref{app:positivity} for detail),
  we have $h_{t\sigma} > 0$ for $q[a-(a^2+r^2)\omega]>0$ and
  $h_{t\sigma} < 0$ for $q[a-(a^2+r^2)\omega]<0$
  at the effective horizon. 
  This indicates that solutions of the $r^+_\mathrm{eff}$ branch have 
  $h_{t\sigma} > 0$ and those of the $r^-_\mathrm{eff}$ branch have
  $h_{t\sigma} < 0$ at the effective horizon.
  The sign of $h_{t\sigma}$ is associated with whether the Killing
  horizon generated by $\xi^a = (\partial_t)^a$ on the world sheet becomes
  black-hole-type for $h_{t\sigma} > 0$ or
  white-hole-type for $h_{t\sigma}<0$. 
  We can simply understand it as follows.
  At the effective horizon, two future-directed null vectors with
  respect to the induced metric are given by 
  \begin{equation}
   \xi^a = \left(\partial_t\right)^a,\quad 
    \chi^a = h_{\sigma\sigma}\left(\partial_t\right)^a- 
    2 h_{t\sigma} \left(\partial_\sigma\right)^a. 
  \end{equation} 
  It turns out that $\xi^a$ is outgoing null vector and $\chi^a$ is
  ingoing one if $h_{t\sigma}>0$, 
  whereas $\xi^a$ is ingoing null vector and $\chi^a$ is outgoing one
  if $h_{t\sigma}<0$.  
  Coordinates on the world sheet can be freely chosen, and 
  dynamics
  of the string dose not explicitly depend on the world sheet
  coordinates, so that a time coordinate on the world sheet has little physical
  meaning in general.
  However, because we have identified the time coordinate on the world sheet with
  the Boyer-Lindquist time $t$ on the spacetime, 
  time evolutions on the world sheet described by this time $t$ have
  physical meanings.
  When $\xi^a$ associated with the physical time $t$ generates the
  white-hole-like effective horizon, 
  we should provide information at the effective horizon
  whenever we solve time evolutions in terms of $t$.
  It means that these configurations will be never realized unless
  specific conditions continue to be provided. 
  On the other hand, when $\xi^a$ generates the black-hole-like effective
  horizon, we do not need any conditions at the effective horizon to
  solve time evolutions because it is located in the causal future.
  Therefore, such configurations can be naturally realized by time
  evolutions without any
  specific initial conditions or any information beyond the stationary region.

\begin{figure}[t]
   \centering
   \includegraphics[width=15cm,clip]{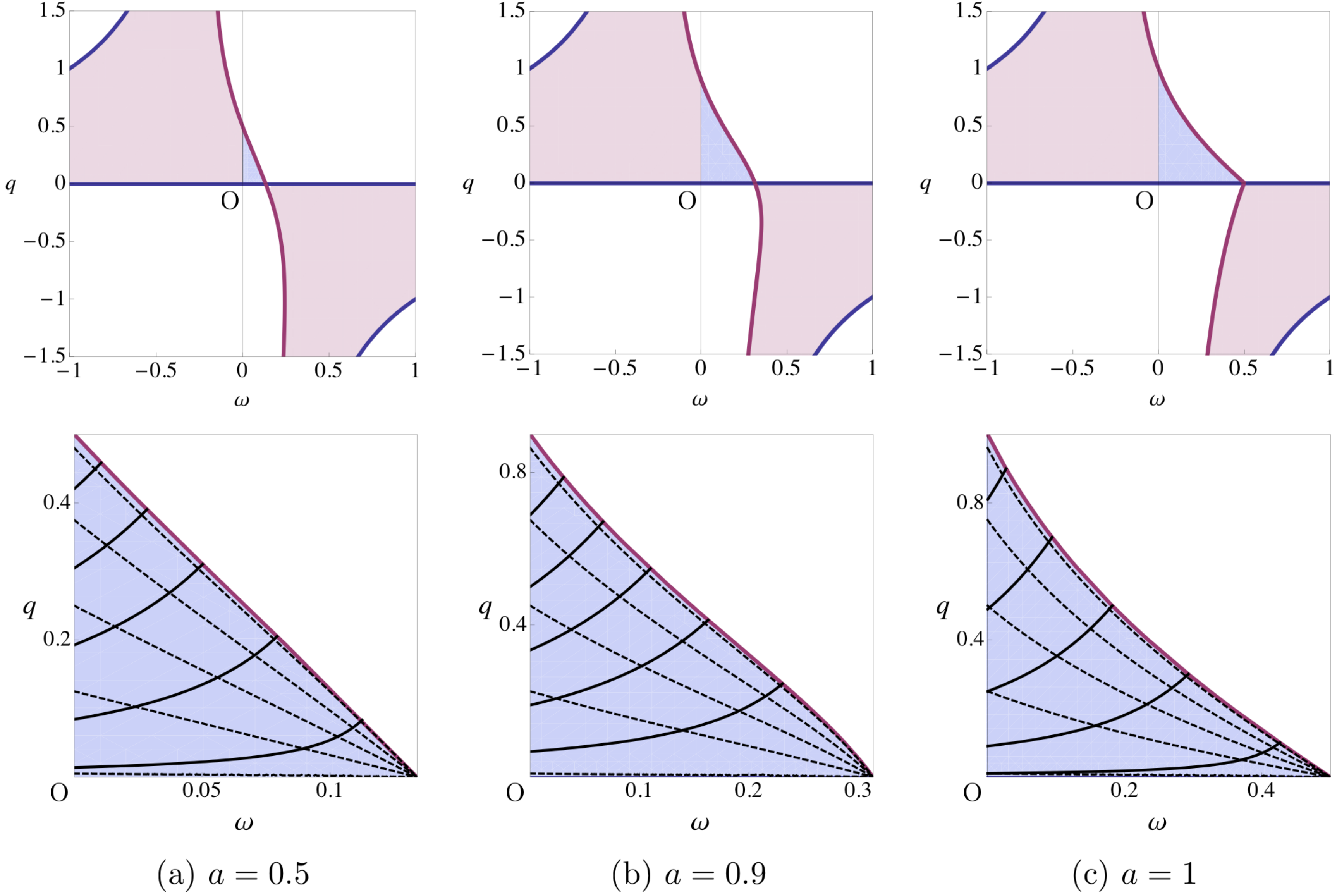}
\caption{
Parameter spaces for the stationary strings 
that pass through the effective horizon regularly 
in the case of (a) $a=0.5$ (left), (b) $a=0.9$ (center), and (c) $a=1.0$ (right) 
   in the unit $M=1$,
   which are shown by the shaded regions in the upper plots. 
Each region is enclosed by $\omega_0(q)$ (blue solid lines) and 
$\omega_{\pi/2}(q)$ (red solid lines). 
The lower plots show contours of $r_{\mathrm{eff}}^+$ and $\theta_{\mathrm{eff}}^+$ 
in the case where positive energy extraction occurs. 
The dashed lines denote the contours of 
$\sin \theta_{\mathrm{eff}}^+=
0.98, \sqrt{3}/2, 1/\sqrt{2}, 1/2, 0.1$ 
from the top to the bottom in each plot. 
The solid gray lines denote the contours of 
(a) $r_{\mathrm{eff}}^+=2, 1.98, 1.95, 1.92, 1.89, 1.87$;  
(b) $r_{\mathrm{eff}}^+=1.9, 1.8, 1.7, 1.6, 1.5, 1.4$; 
(c) $r_{\mathrm{eff}}^+=1.9, 1.7, 1.5, 1.3, 1.1$, 
from the top to the bottom. 
}
\label{fig:parameter_space}
  \end{figure}

\begin{figure}
   \centering
   \includegraphics[clip,width=6cm]{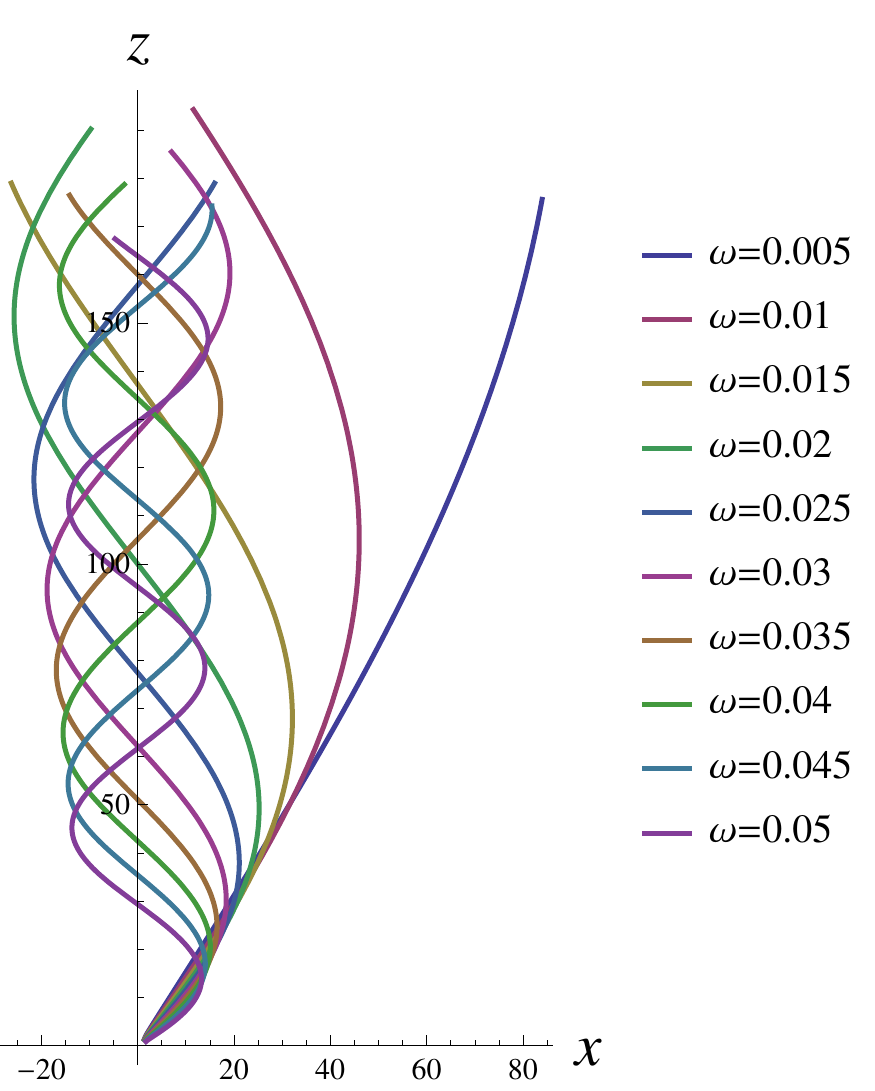}
   \includegraphics[clip,width=8cm]{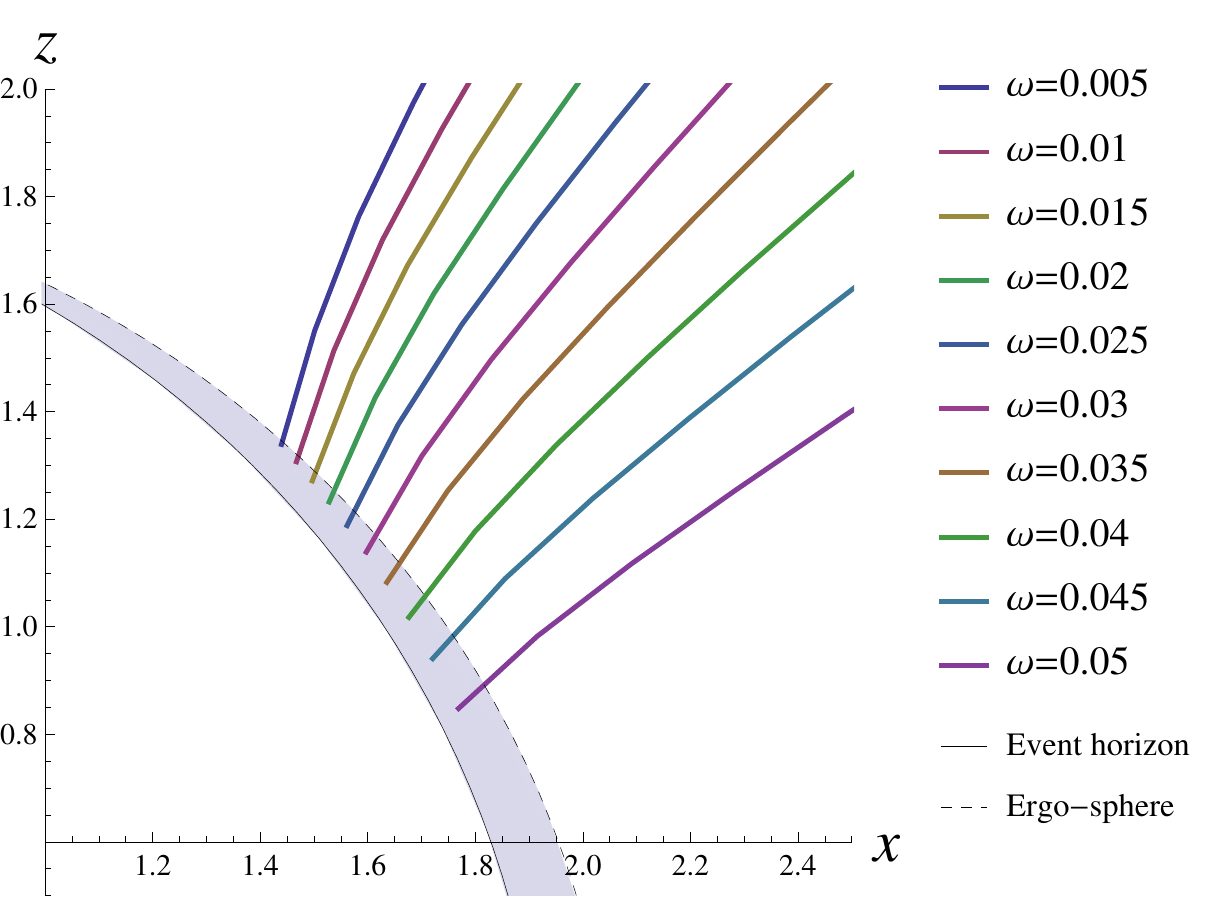}
   \caption{
   Examples of the string configurations which can extract
   positive energy from the black hole for $M=1$, $a=1/2$ and $q=1/4$.
   The left panel shows global configurations of the strings for various
   angular velocities $\omega$.
   As $\omega$ is larger, the pitch along $z$ axis tends to be shorter.
   The right panel shows an enlarged view near the black hole.
   Each end point of the string inside the ergoregion (the shaded
   region) corresponds to the locus of the effective horizon.
   As $\omega$ is larger, the locus of the effective horizon tends to be 
   close to the equatorial plane.
   Coordinates have been taken as
   $(x,z)=(\sqrt{r^2+a^2}
   \sin\theta\sin\phi, r\cos\theta)$. 
   }
   \label{fig:string_configurations}
  \end{figure}

  Parameter spaces for the rigidly rotating 
  strings of the
  $r^+_\mathrm{eff}$ branch are shown in Fig.~\ref{fig:parameter_space}.
  Allowed regions in $(\omega,q)$ space, which is given by
  Eq.~(\ref{eq:allowed_region}), are enclosed by $\omega_0(q)$ and
  $\omega_{\pi/2}(q)$.  
  In the $r^+_\mathrm{eff}$ branch, the 
  energy-extraction process,
  $\omega q >0$, corresponds to the region
  $0<\omega \le \omega_{\pi/2}(q)$ for $q>0$.
  If $q>0$, $r^+_\mathrm{eff}(\omega,q)$ is monotonically decreasing in
  terms of $\omega$ because 
  $\partial r^+_\mathrm{eff}/\partial \omega <0$ (see Appendix~\ref{app:derivation}). 
  As we have mentioned, 
  the effective horizon coincides with the ergosphere
  when $\omega = 0$,
  that is, 
  $r^+_\mathrm{eff}(0,q) = r_\mathrm{ergo}(\theta^+_\mathrm{eff})$.
  It means that $r^+_\mathrm{eff}$ is always less than $r_\mathrm{ergo}$
  in the interval $0<\omega \le \omega_{\pi/2}(q)$ for $q>0$.
  In addition, $\omega_{\pi/2}(q)$ can be rewritten as
  \begin{equation}
   \omega_{\pi/2}(q) = \frac{2Ma}{r^3+a^2r+2Ma^2}
    - q \frac{r}{r^3+a^2r+2Ma^2} ,
  \end{equation}
  where $r$ should satisfy $\Delta(r)=q^2$.
  Since the first term monotonically decreases for $r\ge r_\mathrm{h}$ and
  equals $\Omega_\mathrm{h}$ when $r=r_\mathrm{h}$,
  we have $\omega_{\pi/2}(q) \le \Omega_\mathrm{h}$ for $q\ge 0$ with
  equality if and only if $q=0$.
  As a result, we can conclude that a necessary condition for the energy
  extraction is that the effective horizon on the string should
  enter the inside of the 
  ergoregion and the angular velocity of the
  string should be less than that of the black hole.
  It is obvious that the parameter region where the 
  energy extraction can
  occur becomes wider as the Kerr parameter $a$ is larger, and it
  vanishes for $a=0$.
  In the ($\omega,q$) plane the power of the extraction $\omega q$ is
  described by the area of the rectangle whose sides are $\omega$ axis and
  $q$ axis.
  For a fixed $\omega$, the power will become larger as
  $\theta^+_\mathrm{eff}$ closes to the equatorial plane.
  For a fixed $\theta^+_\mathrm{eff}$, it will be maximized as $\omega$
  becomes about half of the black hole angular velocity $\Omega_\mathrm{h}$. 
  Figure~\ref{fig:string_configurations} shows explicit examples of the
  string configurations in the parameter region where the energy
  extraction occurs.
  We can see that the strings extend sufficiently far from the
  black hole, while the effective horizons of them are located inside
  the ergoregion.

\begin{figure}
   \centering
   \includegraphics[clip,width=5cm]{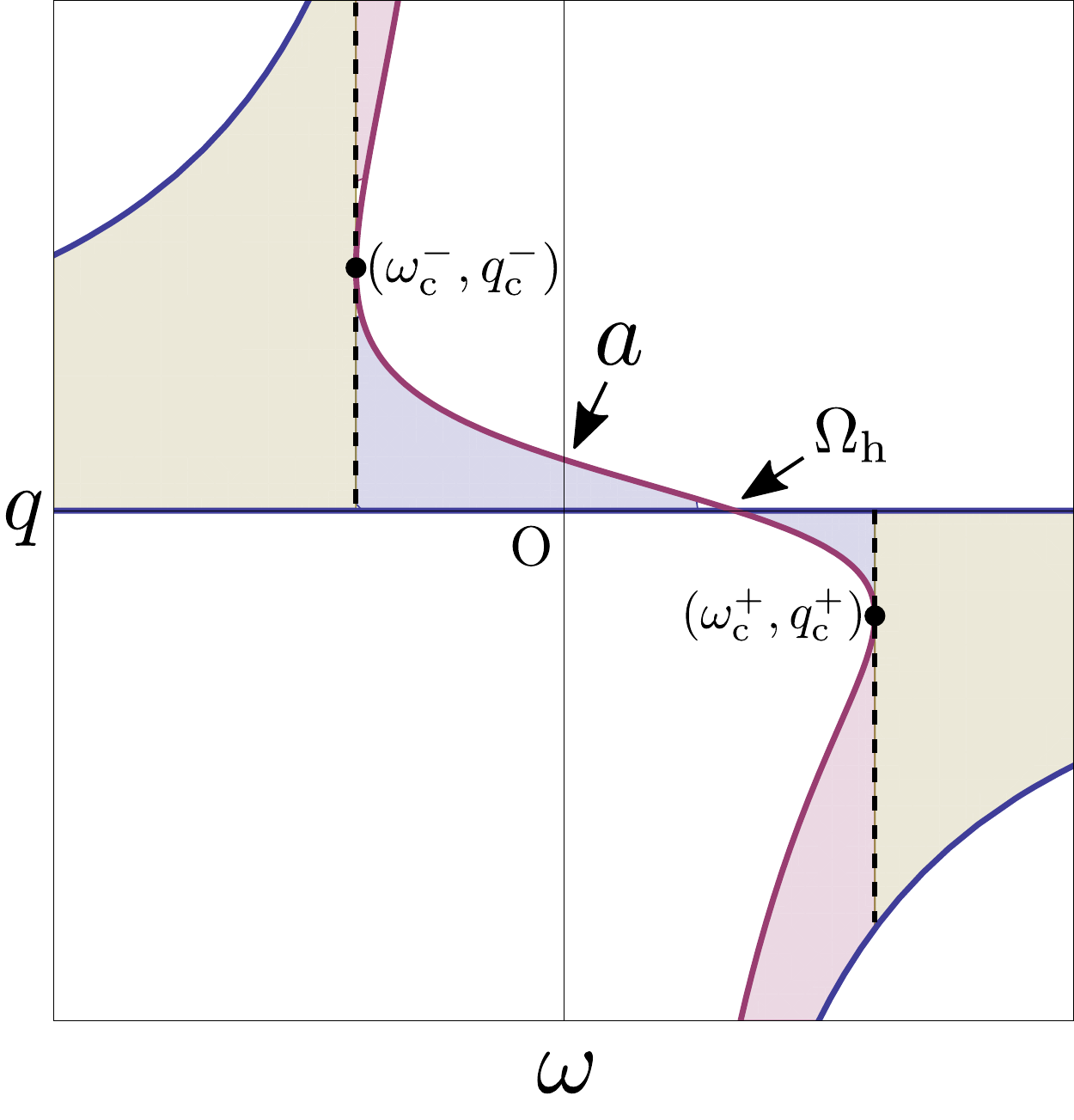}
   \caption{
   Sketch of the parameter region for 
   a typical $a$ ($0<a<1$).
   The effective horizon on the string respectively
   corresponds to (i) the inner
   light surface when
   $\omega^-_\mathrm{c} \le \omega \le \omega^+_\mathrm{c}$ and
   $q^+_\mathrm{c} \le q \le q^-_\mathrm{c}$,
   (ii) the outer light cylinder when 
   $\omega^-_\mathrm{c} \le \omega \le \omega^+_\mathrm{c}$ and
   $q \le q^+_\mathrm{c}, q \ge q^-_\mathrm{c}$, and
   (iii) the connected light surface when
   $\omega \le \omega^-_\mathrm{c} , \omega \ge \omega^+_\mathrm{c}$.
   The critical point $(\omega^\pm_\mathrm{c}, q^\pm_\mathrm{c})$,
   where the three regions join, is the extremum of
   $\omega_{\pi/2}(q)$.
   }
    \label{fig:sketch}
   \end{figure}

  The effective horizon on the induced geometry emerges when the
  rigidly rotating 
  string regularly passes through a light surface on the
  spacetime.
  For a given $\omega$, the effective horizon is uniquely determined,
  whereas there are several possibilities for the light surfaces;  
  the inner light sphere near the black hole and the outer light
  cylinder exist, or the 
  single connected light surface does.
  Therefore, even in the $r^+_\mathrm{eff}$ branch one should be careful
  about which light surface the string is passing through. 
  We find critical angular velocities $\omega_\mathrm{c}^\pm$
  ($\omega^-_\mathrm{c} < 0 < \omega^+_\mathrm{c}$), at which
  the inner and outer light surfaces merge on the equatorial plane, is
  given by the extrema 
  of $\omega_{\pi/2}(q)$, namely
  $\omega^\pm_\mathrm{c} = \omega_{\pi/2}(q^\pm_\mathrm{c})$ such that
  $d\omega_{\pi/2}/dq|_{q=q^\pm_\mathrm{c}}=0$
  (see Appendix~\ref{app:critical} for detail).
  The topology of the light surface changes at $\omega=\omega^\pm_\mathrm{c}$,
  while the configurations of the string may continuously
  deform in terms of the two parameters ($\omega,q$).
  Hence, the parameter region can be classified into the following three
  categories based on the light surface passed by the string: 
  (i) the inner light sphere near the black hole when
  $\omega^-_\mathrm{c} \le \omega \le \omega^+_\mathrm{c}$ and
  $q^+_\mathrm{c} \le q \le q^-_\mathrm{c}$,
  (ii) the outer light cylinder when 
  $\omega^-_\mathrm{c} \le \omega \le \omega^+_\mathrm{c}$ and
  $q \le q^+_\mathrm{c}, q \ge q^-_\mathrm{c}$, and
  (iii) the connected light surface when
  $\omega \le \omega^-_\mathrm{c} , \omega \ge \omega^+_\mathrm{c}$.
  In addition,  
  the points ($\omega^\pm_\mathrm{c}, q^\pm_\mathrm{c}$) are the critical
  points at which the three regions join.
  A sketch of the parameter region is depicted in Fig.~\ref{fig:sketch}.
  It is worth noting that 
  in the region for $\omega q>0$, which we have been interested in, the
  effective horizon on the string is the inner light sphere.
  We can confirm that, when the 
  energy extraction occurs, the string
  is regularly passing through the inner light sphere and 
  is twining around the black hole. 

  This result indicates that the energy extraction can occur when the
  string enters the inside of  
  the ergoregion, and relations between the event horizon and the string
  configuration such as whether the string would intersect with the
  event horizon are irrelevant.
  This is not surprising logically.
  Outward flux conveying energy to infinity, physically, cannot 
  propagate with
  exceeding the speed of light.
  Such energy flux must depend only on its causal past.
  By the definition of the event horizon, the causal past of energy flux
  which can reach to infinity does not include the event horizon.
  Hence, the event horizon is irrelevant to occurrence of the energy
  extraction by outward energy flux.

 \section{Global configurations}
 \label{sec:global}
 
 In this section, we discuss 
 global configurations of the regular string extracting the rotational
 energy from a Kerr black hole in two regimes: 
 One 
 is a slow-rotation regime $a \ll M$ and the other is the extremal
 case $a=M$.
 
  \subsection{Slow-rotation approximation}

For general string configurations with $\omega\neq 0$, we 
do not have analytic solutions of the equations of motion. To study
analytically 
some global
properties of rigidly rotating string configurations, we perform
perturbation analysis of the equations 
of motion by supposing that the
rotation of the Kerr black hole is slow, namely $a\ll M$.
This perturbation analysis can be regarded as
extensions of the work \cite{Frolov:1996xw} to rigidly rotating string
configurations with $\omega\neq 0$ 
in the slow-rotation regime. 
We use the same gauge with 
the one adopted in Ref.~\cite{Frolov:1996xw} 
for
usefulness of comparison with it. Thus the string configuration is given
by the embedding
$t=\tau + \eta(r)$, $\phi =\omega t +\varphi(r)$, 
and $\theta=\theta(r)$.%
\footnote{
Note that $\eta(r)$ means a gauge choice for a time coordinate $\tau$ on
the world sheet, and one can choose $\eta(r)$ so that the
induced metric becomes diagonal in these world sheet coordinates, for example.
However, the following equations of motion by using only quantities
associated with the spacetime
coordinates are independent of the gauge choice on the world sheet.
}
The resultant equations of motion are 
%
\begin{eqnarray}
&&\left(\frac{d\varphi}{dr}\right)^{2} = \frac{Gq^{2}}{\Delta^{2} \sin^{4}{\theta}}, 
\label{eq1}
\\
&&\frac{1}{\sqrt{G}}\frac{d}{dr} \left( \frac{\Sigma F}{\sqrt{G}}\frac{d\theta}{dr} \right) =\frac{\cos{\theta}}{\Delta \sin^{3}{\theta}}Z,
\label{eq2}
\end{eqnarray} 
where
\begin{eqnarray}
&&Z=q^{2}-(q^{2}-\Delta\sin^{2}{\theta})\left( 1-\frac{\Delta(1-a^{2}\omega^{2}\sin^{4}{\theta})}{\Sigma F} \right),
\\
 &&G=\frac{\Sigma F\sin^{2}{\theta}}
 {\Delta\sin^{2}{\theta} - q^{2}}\left[1+\Delta\left(\frac{d\theta}{dr}\right)^{2}\right].
\label{Gdef}
\end{eqnarray} 
Note that 
$\Delta$, $\Sigma$, and $F$ have been 
defined 
in the previous section.
We can reduce Eqs.~\eqref{eq:EOM1affine}--\eqref{eq:EOM3affine} 
to the above equations of motion
with a transformation of variables such as 
$d\theta/dr=\theta'/r'$, $d\varphi/dr=\varphi'/r'$, and so on. 
The solutions of Eqs.~(\ref{eq1}) and (\ref{eq2}) for
$\omega=0$ which are regularly crossing the ergosphere were obtained
analytically in Ref.~\cite{Frolov:1995vp}.
They are explicitly written as 
\begin{equation}
 \theta(r) = \theta_0 ,\quad
  \varphi_\pm(r) = \pm \frac{a}{2\sqrt{M^2 - a^2}}
  \log
  \left(\frac{r-r_\mathrm{h}+2\sqrt{M^2 - a^2}}{r-r_\mathrm{h}}\right) ,
  \label{eq:omega0sol}
\end{equation}
where
$\theta_0$ is the integration constant related to $q$ as 
$q=\pm a\sin^2\theta_0$ 
and
a constant term within $\varphi_\pm(r)$ has been discarded 
thanks to the axisymmetry.

Here we consider string configurations with $\omega\neq 0$ in 
the slow-rotation regime. 
We can define two regions in the Kerr spacetime as
\begin{equation}
 \begin{aligned}
  \text{far region} &: r-r_{\text{h}} \gg M , \\
  \text{near region} &: r-r_{\text{h}} \ll M^{2}/a .
 \end{aligned}
\end{equation}
%
Note that the slow-rotation regime $a \ll M$ allows us to take an overlap
region between the far and near regions as
$M \ll r-r_\mathrm{h} \ll M^2/a$. 
Furthermore we assume the following scalings for $\omega$ and $q$:
%
\begin{eqnarray}
\omega =O(a/M^{2}),\quad q=O(a) ,
\label{setup}
\end{eqnarray}
%
which mean that the strings, also, are slowly rotating. 
In this setting, we can neglect the black hole and regard the spacetime
as a flat spacetime in the far region at the leading order.
At the 
near region, we can solve Eqs.~(\ref{eq1}) and (\ref{eq2}) by
perturbation analysis in 
the $a/M$ expansions.
The outer light cylinder is not located in the near region and the inner
light sphere is not in the far region, so that we should obtain regular
solutions in each region and match the near-region solutions with the
far-region solutions in order to have globally regular solutions.

\subsubsection{Far region}
At the far region the equations of motion reduce to
those in the flat
spacetime. The general solutions of Eqs.~(\ref{eq1}) and (\ref{eq2}) in
the flat spacetime were obtained by 
Ref.~\cite{Frolov:1996xw} 
in cylindrical coordinates as 
%
\begin{eqnarray}
&& z (\rho) 
  = \frac{p}{2\omega}\left( \,\text{arcsin}\frac{B-2\omega^{2}\rho^{2}}{C} +z_{0} \right), 
\label{zsol}
\\
&& \varphi(\rho) =\pm\frac{1}{2}\left(
  \,\text{arcsin}\frac{B\rho^{2}-2q^{2}}{C\rho^{2}}
  +\omega q\,\text{arcsin}
  \frac{B-2\omega^{2}\rho^{2}}{C}
  +\varphi_{0}
\right),
\label{vsol}
\end{eqnarray} 
where 
\begin{eqnarray}
B\,\equiv\, 
1-p^{2}+q^{2}\omega^{2},~~
C\,\equiv\, 
\sqrt{B^{2}-4\omega^{2}q^{2}}, 
\end{eqnarray} 
and $z_{0}$, $\varphi_{0}$, and $p$ 
are integration constants. 
The
coordinates $z$ and $\rho$ can be identified with $z=r\cos{\theta}$ and
$\rho=\sqrt{r^2 + a^2}\sin{\theta}$
in the Boyer-Lindquist coordinates, respectively. 
The constant $p$
corresponds to the momentum flux per unit length along the cylinder, and
the constants $z_{0}$ and $\varphi_{0}$ are 
appropriate offsets for the positions.
They should be determined by matching with near-region solutions
later. 
The string in the flat spacetime can have only one
effective horizon, which we identify with the outer light cylinder, at
$\rho=\omega^{-1}$ since $F=1-\omega^{2}\rho^{2}$. The string
configuration in the flat spacetime given by the 
solutions (\ref{zsol}) and
(\ref{vsol}) is constrained to lie in
$\rho_{-}\le\rho\le\rho_{+}$,  
where $\rho_{\pm}$ defined by 
%
\begin{eqnarray}
 \rho_{\pm}
 \,\equiv\, 
 \frac{1}{\omega}\sqrt{\frac{B\pm C}{2}}
  \label{eq:turning_points_flat}
\end{eqnarray}
%
are turning points of string configurations.%
\footnote{ This constraint is equivalent to the 
constraints for the
argument of $\arcsin$ in Eqs.~(\ref{zsol}) and (\ref{vsol}), 
namely
$-1 \le \left(B-2\omega^{2}\rho^{2}\right)/C \le 1$ 
 and
$-1 \le \left(B\rho^2 - 2q^2\right)/\left(C\rho^2\right) \le 1$.
}
If we require the regularity condition on the outer light cylinder, we
have $\rho_{+}=\rho_{-}=\omega^{-1}$ with $q\omega=1$.
Such regularity condition constraints the configuration of regular strings completely on the outer light cylinder, and it cannot be extended from the outer light cylinder. We are now interested in the string configurations 
which extend into the inside of the outer light cylinder and reach the inner light sphere to extract the rotation energy of the Kerr black hole. 
Therefore, we 
will not impose the regularity condition at the outer light cylinder, and instead, 
consider solutions satisfying
%
\begin{eqnarray}
 \rho_{+}<\omega^{-1} .
  \label{pcond}
\end{eqnarray}
%
Such solutions never reach the outer light cylinder, 
and their string configurations can be regular 
without satisfying the regularity condition on the outer light cylinder.   
The condition (\ref{pcond}) is regarded as that for $p$. 
As we will see below, $p$ obtained by matching actually satisfies the condition.

\subsubsection{Near region}

Next, in the near region, we perturbatively solve the equations of
motion~\eqref{eq1} and \eqref{eq2} in the $a/M$ expansions.
With $a=0$ and $\omega =0$ the solution of Eqs.~(\ref{eq1}) and
(\ref{eq2}) is given by 
%
\begin{eqnarray}
\theta(r)=\theta_{0}, \quad \varphi(r)=0,
\end{eqnarray}
%
which is derived from Eq.~\eqref{eq:omega0sol} with $a=0$ and leads to
$q=0$.
Assuming this solution as a zeroth-order solution, we expand
$\varphi(r)$, $\theta(r)$, $q$, and $\omega$ as
\begin{eqnarray}
 \theta(r) = \theta_{0} +\theta_{2}(r)\left(\frac{a}{M}\right)^{2}
  +\theta_{4}(r)\left(\frac{a}{M}\right)^{4}+\cdots, \quad
  \varphi(r) = \varphi_{1}(r)\left(\frac{a}{M}\right)
  +\varphi_{3}(r)\left(\frac{a}{M}\right)^{3}+\cdots ,
\end{eqnarray}
\begin{eqnarray}
 q=q_{1}\left(\frac{a}{M}\right)
  +q_{3}\left(\frac{a}{M}\right)^{3}+\cdots, \quad
 \omega=\omega_{1}\left(\frac{a}{M}\right)
 +\omega_{3}\left(\frac{a}{M}\right)^{3}+\cdots .
\end{eqnarray}
We solve the equations of motion order by order under these
expansions 
with the regularity condition that the solution is regular at the inner
light sphere $r=r^{\pm}_{\text{eff}}$. 
As we have seen, 
this regularity condition gives the boundary conditions for string configurations at the inner light sphere. 
Expanding $r^{\pm}_{\text{eff}}$ in terms of $a/M$ yields 
%
\begin{equation}
 r^\pm_\mathrm{eff} = 2M
  - [M \mp q_1 (1 - 4M\omega_1)]\frac{a^2}{2M^2}
  +O(a^{4}/M^{3}) .
  \label{rmeff}
\end{equation}
%
Thus the regularity condition, for example, requires that
$\theta_{2}(r)$, $\theta_{4}(r)$ and so on 
should be regular at $r=2M$ in the
$a/M$ expansions. 
In contrast, since the positions of the inner light surface and the event horizon are
degenerate at the zeroth order in the $a/M$ expansions, the function 
$\varphi_{1}(r)$ can be singular in logarithm at $r=2M$ in the
Boyer-Lindquist coordinates due to the rotation effect of the black hole.
Thus we need not impose the regularity of the functions $\varphi_{1}(r)$,
$\varphi_{3}(r)$ and so on, at $r=2M$. 
Then, by solving Eqs.~(\ref{eq1}) and (\ref{eq2}) with the regularity
condition in the $a/M$ expansions, at the leading order we have regular
solutions in the near region as 
%
\begin{eqnarray}
 &&\varphi^{\pm}_{1}(r)
  = \varphi_{1}^{0} - \frac{q_{1}^\pm\left[\log{(r-2M)}-\log{r}\right]}
  {2M\sin^2\theta_0},
\label{phisol}
\\
&&
\theta_{2}(r)=\theta_{2}^{0} -\frac{5Mr\omega^{2}_{1}\sin{2\theta_{0}}}{6}-\frac{r^{2}\omega_{1}^{2}\sin{2\theta_{0}}}{12} 
-\frac{11M^{2}\omega_{1}^{2}\sin{2\theta_{0}}}{3}\log{\frac{r}{2M}}\notag \\
&&~~~~~~~~~
-M\omega_{1}(1-2M\omega_{1})\sin{2\theta_{0}}\left[
\text{Li}_{2}\left(1-r/2M\right)
+
\frac{1}{2} \left(\log{\frac{r}{2M}}\right)^{2}\right],
\label{thesol}
\end{eqnarray} 
where $\varphi_{1}^{0}$ and $\theta_{2}^{0}$ are integration constants
and $\text{Li}_{2}(x)$ denotes the polylogarithm function.
For simplicity, we will omit the constant terms
$\varphi_{1}^{0}$ and $\theta_{2}^{0}$ hereafter,
because they can be absorbed by
the constant terms in the zeroth-order solutions. 
The second and third terms 
in the right-hand side 
of Eq.~(\ref{thesol}) imply the breakdown of the near-region solution at $r=O(M^{2}/a)$. 
The regularity condition for $\theta_{2}(r)$ at
the inner light sphere requires
%
\begin{eqnarray}
q_{1} =q_{1}^{\pm} \equiv \pm M(1-4\omega_{1}M)\sin^{2}{\theta_{0}}.
\label{q1cond}
\end{eqnarray}
%
In Eq.~(\ref{thesol}) we have already used this condition.
We note that the above condition is identical to the regularity
condition $q^2=\Delta \sin^2\theta|_{r=r^\pm_\mathrm{eff}}$ of Eq.~(\ref{eq:regularity_condition}) at the effective horizon in
the $a/M$ expansions.

Let us confirm some properties shown in the previous section for
the near-region solution under the slow-rotation approximation.
The regular near-region solutions
with $q_{1}=q_{1}^{\pm}$ correspond to
the $r^{\pm}_{\mathrm{eff}}$ branches.
The condition (\ref{q1cond}) yields 
%
\begin{eqnarray}
q = \pm 4M^{2}\left( \Omega_{\text{h}} -\omega
					       \right)\sin^{2}{\theta_{0}}
+ O(a^{3}/M^{2}) ,
\end{eqnarray}
where we have used $\Omega_\mathrm{h} = a/(4M^2) + O(a^3/M^4)$.
For $\omega<\Omega_{\mathrm{h}}$, the $r^{+}_{\mathrm{eff}}$ branch has a
positive $q$ and the $r^{-}_{\text{eff}}$ branch has a negative $q$. 
As we have seen, the physically reasonable solution is given by 
$q_{1}=q_{1}^{+}$, and $q_{1}=q_{1}^{-}$ gives its time-reversal 
solution. Indeed we can see that the physically reasonable solution has the positive energy flux at the inner light surface only when the inner light surface is located in the ergoregion. The energy flux at the inner light surface is given by
%
\begin{eqnarray}
\omega q &=& \frac{\omega_{1}a^{2}}{M}\left(1-4M\omega_{1}\right)\sin^{2}{\theta_{0}}
+O(a^{4}/M^{4}) 
\notag \\
&=& 4M^{2}\omega(\Omega_{\text{h}}-\omega)\sin^{2}{\theta_{0}} 
+O(a^{4}/M^{4}).  
\end{eqnarray}
The locus of the ergosphere is 
%
\begin{eqnarray}
 r_{\text{ergo}} (\theta_0) 
  &=& 2M -\frac{a^{2}\cos^{2}{\theta_{0}}}{2M}+O(a^{3}/M^{2}).
\end{eqnarray}
%
Using Eq.~(\ref{q1cond}) we have
%
\begin{equation}
 \begin{aligned}
  r_{\mathrm{ergo}}(\theta_0) -r^{+}_{\mathrm{eff}}
  &= 4a^2\omega_1 (1-2M\omega_{1})\sin^{2}{\theta_{0}}+O(a^{3}/M^{2}) \\
  &= 
  8M^3\omega \left(2\Omega_\mathrm{h} - \omega\right) \sin^{2}{\theta_{0}}
  + O(a^{3}/M^{2}) .
 \end{aligned}
\end{equation}
%
Thus the positive energy flux is realized only if
 $r_{\text{ergo}}>r^{+}_{\text{eff}}$.
 Note that the solutions in the
 $r^+_\mathrm{eff}$ branch 
 with $2>4M\omega_{1}>1$ have the negative energy flux
 although their inner light surfaces are inside the ergoregion. This is
 because such solutions are rotating faster than the Kerr black hole and
 supplying the angular momentum to the black hole.
 One
 interesting observation on the energy flux is the fact that the maximum
 energy flux is realized when the angular velocity of the string is half
 of the horizon angular velocity  as  $\omega=\Omega_{\text{h}}/2$
 for a given $\theta_0$.
 This is the same situation with 
 the Blandford-Znajek process for force-free magnetosphere \cite{Blandford:1977ds}.

\subsubsection{Matching}

Let us perform the matching of the far-region and near-region solutions to see the
global configuration of the solution. The behavior of the far-region
solution in the overlap region, $M \ll r-r_\mathrm{h} \ll M^2/a$,  
can be obtained by using the $a/M$ expansions to the far-region solution
(\ref{zsol}) and (\ref{vsol}).
Then we find that the conditions for the 
matching are 
%
\begin{eqnarray}
 p=\cos{\theta_{0}} +O(a/M),
  \quad
  z_0 = \frac{\pi}{2}+ O(a/M),\quad
  \varphi_0 = - \frac{\pi}{2} + O(a/M) .
\end{eqnarray}
%
Under these conditions we can match the regular near-region solution with
the far-region solution consistently. Furthermore, for this value of
$p$,
the outer turning point 
$\rho_{+}$ satisfies the condition (\ref{pcond})
%
\begin{eqnarray}
\rho_{+} = \frac{M\sin{\theta_{0}}}{a\omega_{1}} <\frac{M}{a\omega_{1}} 
\end{eqnarray}
%
except for the string on the equatorial plane $\theta_{0}=\pi/2$.
This means that the solution with $0<\theta_{0}<\pi/2$ can start from the
inner light sphere which is located inside the ergoregion, and extend
to 
the infinity without reaching the outer light cylinder in the
slow-rotation limit.
As a result, we can extract the rotational energy of the Kerr black hole
to 
the infinity 
by the rigidly rotating regular strings.
This is the
result just obtained by perturbation analysis in the slow-rotation
approximation.
However, we will see that 
similar properties hold also even in not
slow-rotation regime below, and it means that
the extraction of the rotational energy 
from the Kerr black hole without touching the outer light
cylinder by the rigidly rotating regular strings can generally occur. 

\subsection{The extremal Kerr background}

We focus on the global configuration of 
the stationary rotating string 
extracting the rotational energy from the extremal Kerr black hole. 
As shown in Sec.~\ref{sec:rigidly_rotating_string}, 
since 
the area in $(\omega, q)$ space in which 
the string extracts positive energy 
becomes maximum in $a=M$, 
then we expect to see the difference of 
the string configuration obviously 
due to the choice of the parameters.

\begin{figure}[t]
\centering
 \includegraphics[width=17.0cm,clip]{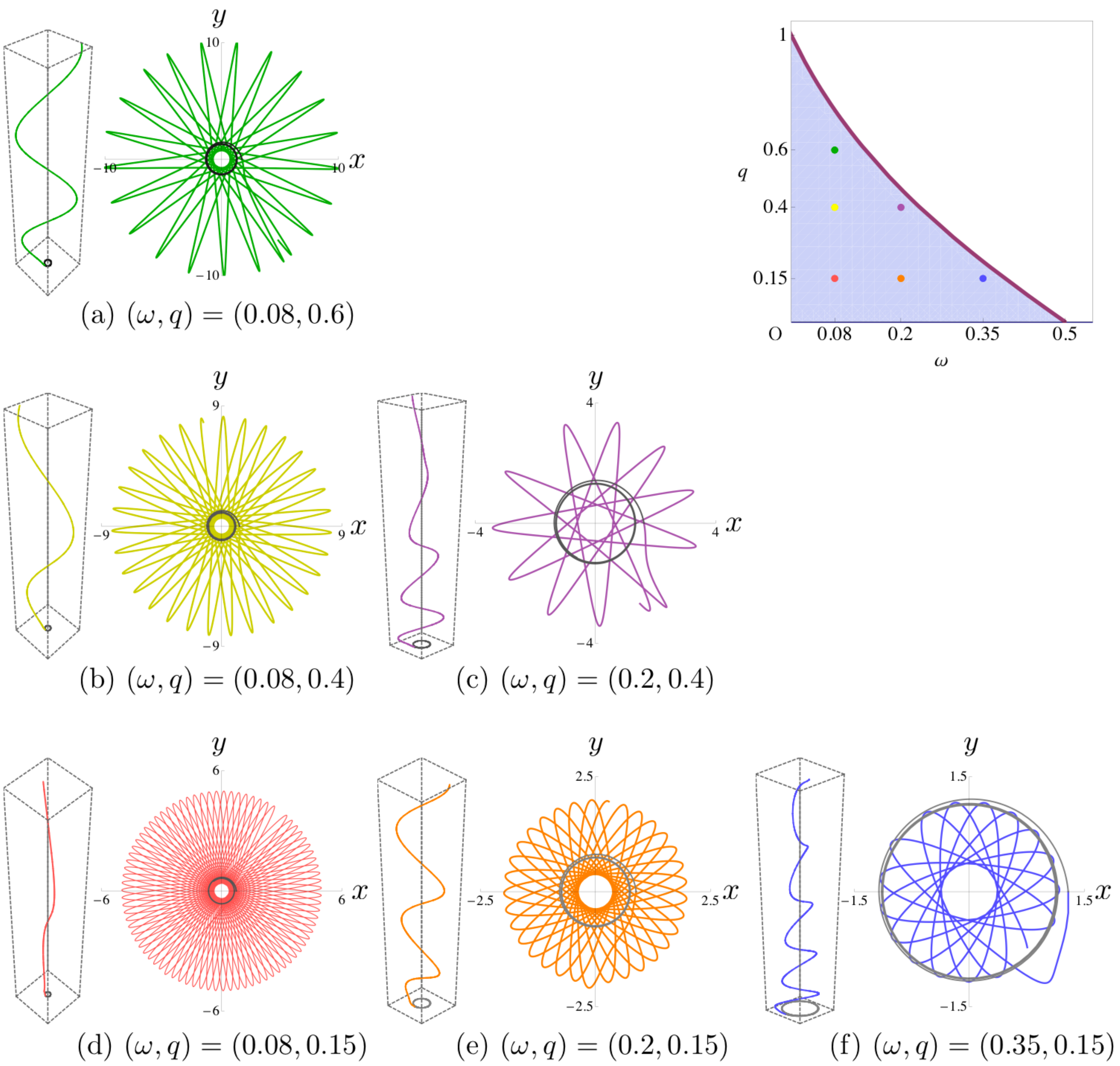}
 \caption{
 Snapshots of a stationary rotating string twining around the extremal Kerr black hole, $a=1$, 
 in units where $M=1$. 
 The left in each figure is a three-dimensional snapshot of the string at constant-$t$ slice, 
 where the plot range is $0\leq z\leq70$ from the bottom to the top. 
 The right in each figure is 
 the string configuration projected onto a constant-$z$ slice. 
 The solid lines colored with gray show the string configuration 
 in the inside of the effective horizon, which are
 twining around the event horizon. 
 The solid lines colored with green, yellow, purple, red, orange, and blue 
 show the string configuration in the outside of the 
 effective horizon. 
 The plot in the upper right corner gives the parameter region of $(\omega, q)$ 
 with which positive energy extraction by the string occurs, 
 where each colored point identified with that of each string
 indicates to the parameter value in panels~(a)-(f).
 Moreover, the arrangement of the points in the parameter region is
 identified with that of the panels~(a)-(f).
 }
 \label{fig:globalconfig}
\bigskip
\end{figure}

Figure~\ref{fig:globalconfig} shows the 
global configurations of 
the stationary rotating string on a
constant-$t$ slice that carries 
positive energy from the extremal black hole. 
Inside the corresponding parameter region, 
we have selected the six sets of the values 
$(\omega M, q/M)
=(0.2, 0.4)$, 
$(0.35, 0.15)$,
$(0.08, 0.6)$, 
$(0.08, 0.4)$, 
$(0.2, 0.15)$, and 
$(0.08, 0.15)$, 
which are in order of decreasing the amount of the energy flux. 
Note that 
the string with $(\omega M, q/M)=(0.2, 0.4)$ 
is the closest to 
the string with the maximum efficiency of the energy extraction. 
The string in each figure crosses over the effective horizon at 
$(r, \theta, \phi)=(r_\textrm{eff}^+, \theta_\textrm{eff}^+, 0)$, 
which is shown as the connection of two colored lines, 
and 
extends from the event horizon to a far region. 
We can conclude that for each set of the parameters the string is
twining around the black hole%
\footnote{ On constant-$t$ slices in the Boyer-Lindquist coordinates the
string seems to be twining around the event horizon endlessly without
crossing it, that is, $\varphi(\sigma)$ will diverge as $r(\sigma)$
goes to $r_\mathrm{h}$.
However, it does not mean that the string can never penetrate the event
horizon because these time slices intersect with the event horizon only
at the bifurcation surface.
In fact, the logarithmic divergences which have appeared in
the exact solution (\ref{eq:omega0sol}) and  
the approximate solution (\ref{phisol}) for $\varphi(\sigma)$ originate
from the regular behavior of $\Delta d\varphi/dr$ at $r=r_\mathrm{h}$,
so that the string can be across the event horizon on other
time slices in horizon-penetrating coordinates such as the
Eddington-Finkelstein coordinates.
In general, we expect that the physically reasonable solutions in the
$r^+_\mathrm{eff}$ branch can naturally penetrate the (future) event
horizon while they cannot be across the (past) white hole horizon. }
and extending to 
the infinity along the
rotational axis when the energy extraction occurs.

Let us introduce the cylindrical coordinates~$(\rho, \phi, z)$ 
that are defined as 
$\rho=\sqrt{r^2+a^2}\sin \theta$. 
The global configuration of the string apart from the black hole is characterized by three
characteristics: 
the ``size'' in $\rho$ direction; 
the ``pitch'' in $z$ direction; 
and the ``frequency'' of $\rho(\sigma)$ in terms of $\phi$, 
which we have defined as 
the proper length in $\rho$ direction at the maximum value of $\rho(\sigma)$; 
the proper length in $z$ direction during one period of the oscillation of $\rho(\sigma)$; 
and 
the number of the oscillation of $\rho(\sigma)$ during one period of the 
oscillation of $\phi(\sigma)$, respectively.

These characteristics are related to the parameters $(\omega, q)$ as follows. 
As seen in Fig.~\ref{fig:globalconfig}, 
the size becomes smaller with increasing $\omega$ or decreasing $q$. 
The pitch in $z$ direction becomes shorter with increasing $\omega$. 
The frequency of $\rho(\sigma)$ in terms of $\phi$ 
decreases as $\omega$ or $q$ increases.

When the string goes apart from the black hole, dynamics of the string
seems to be well described by that in the flat spacetime as well as in
the slow-rotation case.
In fact, the behaviors of the string in Fig.~\ref{fig:globalconfig} are
similar to those in the flat spacetime in Ref.~\cite{Ogawa:2008qn}.
Let us examine the above characteristics by using the general solutions of the rigidly rotating 
string in
the flat spacetime, which have been already shown in Eqs.~(\ref{zsol})
and (\ref{vsol}).
Since the 
corotating 
Killing vector exists among the Kerr and flat
spacetime, $\omega$ and $q$ in the parameters characterizing the
solutions can be identified in the both spacetimes.
However, the translational vector along the $z$ direction is no longer a
Killing vector in the Kerr spacetime, so that a quantity corresponding
to $p$ cannot be conserved.
As seen previously, $p$ in a far region should be determined by
solving the equations of motion near the black hole.
Because the rigidly rotating 
strings have the energy flux
$\omega q \lesssim 0.15$ at most when the energy extraction occurs, we will
read some characteristic quantities from the exact
solutions in the flat spacetime assuming $\omega q$ is small.
The size given by $\rho_+$ of Eq.~(\ref{eq:turning_points_flat}) 
becomes
$\rho_+ \simeq \sqrt{1-p^2}/\omega$ and 
the pitch given by the factor in the front of the first term of
Eq.~(\ref{zsol}) 
becomes $p/\omega$.
Since the factor of the second term of Eq.~(\ref{vsol}) 
represents a
phase shift, the frequency becomes $\sim (\omega q)^{-1}$.
If we suppose $p \simeq \cos \theta^+_\mathrm{eff}$ following the result
in the slow-rotation approximation, the behaviors of the characteristics observed in Fig.~\ref{fig:globalconfig} seem to be
explained.

 \section{Summary and Discussion}

 We have studied rigidly rotating 
 Nambu-Goto strings in the Kerr
 spacetime and 
 have shown that the rotational energy of the black hole
 can be 
 extracted by the strings.
 The string configurations are characterized by two parameters: the
 angular velocity $\omega$ and the angular momentum flux $q$.
 We have considered the string regularly passing through a light
 surface and 
 have analytically exhibited the allowed region where such
 strings exist in the parameter space ($\omega,q$).
 We have found a necessary condition for the energy extraction is that
 the effective horizon on the world sheet caused by the rigid rotation
 will enter into the ergoregion of the Kerr black hole and the angular
 velocity of the rigid rotation is less than that of the black hole,
 namely $\omega<\Omega_\mathrm{h}$.
 Moreover, global configurations of such strings have been examined in a
 slow-rotating case ($a \ll M$) and the extremal case ($a=M$).
 In the both cases we have shown the rigidly rotating 
 strings with
 positive energy flux can start from the inner light sphere inside the
 ergoregion and extend to 
 the infinity.
 It turns out that the energy extraction from the black hole can
 generally occur.

 The current mechanism of the energy extraction can be briefly
 interpreted on the basis of the usual Penrose process in the
 ergoregion
 (the analogy of the Penrose process was mentioned in the
 literature~\cite{Kuiroukidis:1999bc}).
 As we have mentioned, the effective horizon on the world sheet corresponds to
 a stationary limit surface with respect to the 
 corotating 
 Killing vector with angular velocity $\omega$.
 Beyond the effective horizon on the world sheet of the rigidly rotating
 string, the Killing vector tangential to the world sheet becomes spacelike.
 This does not mean that the proper motion of line elements of the
 string may become superluminal.
 The line elements cannot follow the superluminal Killing orbit and
 the interval of the line elements continues to be larger, that is, 
 the string is not stationary but stretching in this region.
 In general, if a string with a tension is stretching, its potential
 energy will increase and the string should consume an energy to stretch.
 However, the situation changes in the ergoregion.
 Because the Killing energy can be negative in the ergoregion,
 the string stretching can decrease its energy similar
 to the fragmentation in the Penrose process for particles.
 Hence, if the effective horizon on the string enters into the
 ergoregion, the string can gain an energy by stretching in the
 ergoregion and extract the energy to 
 the infinity.
 This extraction mechanism is quite simple and general.
 For various other stringlike objects as well as Nambu-Goto strings,
 it is expected that these mechanisms of the energy extraction do work well. 

 It should be emphasized that the necessary condition of the
 energy extraction is determined locally near the effective horizon on a
 light surface but irrelevant of global nature such as configurations of
 the string at
 the event horizon or 
 the infinity.
 This result is reasonable in the following respects.
 This energy-extraction mechanism can be regarded as a complex of
 different processes spatially and temporally separated;
 ``generation'' of an energy in the ergoregion,
 ``transport'' of the energy to 
 the infinity, and
 ``disposal'' of residues resulting from the energy generation.
 Precisely speaking, our necessary condition is a condition for
 generation of the energy.
 In contrast, existence of event horizons is helpful for the
 ``disposal'' process because event horizons of black holes are certainly ideal
 disposal sites, but not so significant for the ``generation'' process.
 Furthermore, in realistic situations of the energy extraction from
 black holes only a single object or phenomenon does not need to play a
 major role in all of the above processes.
 For example, even though a rigidly rotating 
 string cannot entirely reach 
 the
 infinity, the energy extraction will be successful as long as the string
 can reach sufficiently far from the black hole and then transfer its
 energy to other objects without falling back to the black hole.
 As a result, what is most essential and primal in the energy-extraction mechanism is
 the generation process in the ergoregion, and we expect this fact
 may be true for various mechanisms of extracting the rotational energy
 of black holes other than strings discussed in this paper.

 Finally, let us quantitatively evaluate the energy-extraction rate
 $dE/dt = \mu q\omega$ for the string with a tension $\mu$.
 It is worth noting that this energy-extraction rate is irrelevant to the
 mass scale of the central black hole, which is canceled out, even
 though the mass scale determines the amount of the rotational energy to
 extract. 
 Thus, the string tension dominates the energy-extraction rate via this
 mechanism.
 We shall restore the fundamental constants $G_\mathrm{N}$ and $c$, and
 then we have the energy-extraction rate as 
 \begin{equation}
  \frac{dE}{dt} 
   = 5.8 \times 10^{51}\text{erg/s}
   \left(\frac{G_\mathrm{N}\mu/c^2}{1.3 \times 10^{-7}}\right)
   \left(\frac{q/ac^2}{1/2}\right)
   \left(\frac{\omega/\Omega_\mathrm{h}}{1/2}\right) u(\alpha) ,  
 \end{equation}
 where $u(\alpha) \equiv 1 - \sqrt{1-\alpha^2}$ ($0\le u(\alpha) \le 1$) 
 and $\alpha \equiv a c^2/G_\mathrm{N}M$ is the dimensionless Kerr
 parameter.%
 \footnote{ We have assumed the dimension of $\mu$ is
 $\mathrm{M}\mathrm{L}^{-1}$ and the
 dimension of $a$ is $\mathrm{L}$, where $\mathrm{M}$ and $\mathrm{L}$
 respectively denote mass and
 length. }
 The string tension $\mu c^2$ has been normalized by an observational
 upper bound on the cosmic string tension~\cite{Ade:2013xla}.
 Furthermore, it is interesting and suggestive to apply magnetospheres
 around a rotating black hole.
 We consider there are magnetic fields $B$ surrounding a rotating black hole.
 Total magnetic tension $\mu_\mathrm{B}c^2$ is roughly estimated by 
 $\mu_\mathrm{B}c^2 = 4\pi r_\mathrm{h}^2 B^2/\mu_0$,
 where $\mu_0$ is the vacuum permeability.
 Suppose the magnetic field $B \simeq 10^{15} \text{G}$ and
 the black hole mass $M \simeq 10 M_{\odot}$, we have
 $G_\mathrm{N}\mu_\mathrm{B}/c^2 \simeq 10^{-7}$.
 This tension gives a maximum energy-extraction rate $\sim 10^{51}\text{erg/s}$ as
 we have estimated, and this value is comparable to the usual power caused
 by the Blandford-Znajek process for the same magnetic field and black
 hole mass.
 Thus we expect that the magnetic tensions rather than the magnetic
 pressure may play an essential role in the
 energy extraction by the Blandford-Znajek process in black hole
 magnetosphere.

 \begin{acknowledgments}
  We would like to thank Chulmoon~Yoo and Tsuyoshi~Houri
  for helpful discussions.
  This work was supported by JSPS KAKENHI Grants No.~JP16K17704~(S.~K.), No.~JP14J03387~(K.~T.) and MEXT-Supported Program for the Strategic Research Foundation at Private Universities, 2014-2017~(T.~I.). 
 \end{acknowledgments}
  
 \appendix

 \section{DERIVATIONS OF VARIOUS QUANTITIES}

  \subsection{Locus of the effective horizon}
  \label{app:derivation}
  
 As we have 
 seen in Sec.~\ref{sec:rigidly_rotating_string}, 
 for given $\omega$ and $q$, 
 the conditions for the 
 effective Killing horizon are given by 
 \begin{equation}
  \Delta(r) = \frac{[a - (a^2+r^2)\omega]^2}
   {(1 - a\omega \sin^2\theta)^2}\sin^2\theta ,
   \label{eq:condition1_app}
 \end{equation}
 and the regularity condition 
 \begin{equation}
  \Delta(r) \sin^2\theta = q^2 ,
   \label{eq:condition2_app}
 \end{equation}
 in Eqs.~(\ref{eq:horizon_condition2}) and (\ref{eq:regularity_condition}).
 If $\Delta(r_\mathrm{eff}) = 0$, that is, 
 $r_\mathrm{eff} = r_\mathrm{h}$,
 we have $q=0$ immediately.
 It results in $\theta_\mathrm{eff} = 0$ and arbitrary $\omega$, or
 $\omega = \Omega_\mathrm{h}$ and arbitrary $\theta_\mathrm{eff}$.
 Hereafter we assume that $\Delta>0$, namely 
 the effective horizon is located outside the event horizon
 $r_\mathrm{eff}>r_\mathrm{h}$.
 Eliminating $\sin^2\theta$ from the above equations, we have
 \begin{equation}
  q^2 (\Delta/q^2 - a\omega)^2 = [a-(a^2+r^2)\omega]^2 .
   \label{eq:quartic_equation}
 \end{equation}
 This is a quartic equation in terms of $r$ and its positive
 roots will give the radius of the effective horizon.
 If $q [a-(a^2+r^2)\omega] > 0$, then two roots of
 Eq.~(\ref{eq:quartic_equation}) 
 are 
 \begin{equation}
  r = \frac{M}{1+q\omega} \pm 
   \sqrt{\left(\frac{M}{1+q\omega}\right)^2 - a(a-q)} ,
 \end{equation}
 If $q [a-(a^2+r^2)\omega] < 0$, then two roots of
 Eq.~(\ref{eq:quartic_equation}) 
 are 
 \begin{equation}
  r = \frac{M}{1-q\omega} \pm
   \sqrt{\left(\frac{M}{1-q\omega}\right)^2 - a(a+q)} .
 \end{equation}
 Requiring the conditions (\ref{eq:condition2_app})
 and
 $0 \le \sin^2\theta \le 1$ we can exclude the roots with the lower sign 
 among the above four
 roots, as we will see later,
 so that we have the radius of the effective horizon
 \begin{equation}
  r^\pm_\mathrm{eff} \equiv \frac{M}{1 \pm q\omega} +
   \sqrt{\left(\frac{M}{1 \pm q\omega}\right)^2 - a(a \mp q)} .
 \end{equation}
 For each blanch 
 $r=r_\mathrm{eff}^\pm$, 
  the polar angle of the effective horizon 
  $\theta_\mathrm{eff}^\pm$ is
 given by
 \begin{equation}
  \sin^2\theta^\pm_\mathrm{eff} \equiv \frac{q^2}{\Delta(r^\pm_\mathrm{eff})} . 
 \end{equation}
 As an important property of $r^\pm_\mathrm{eff}(\omega,q)$, we have 
 \begin{equation}
  \frac{\partial r^\pm_\mathrm{eff}}{\partial \omega}(\omega,q)
   = \mp q \frac{Mr^\pm_\mathrm{eff}}{(1\pm q\omega)^2} 
   \left[
    \left(\frac{M}{1\pm q\omega}\right)^2 - a(a\mp q)
   \right]^{-1/2} ,
 \end{equation}
 in which $q$ only can change its sign.
 Thus, it turns out that $r^\pm_\mathrm{eff}$ is a monotonic function of
 $\omega$ for a fixed $q$.

 We show the exclusion of the roots $r=r_\mathrm{ex}$, where 
 \begin{equation}
  r_\mathrm{ex} \equiv \frac{M}{1 \pm q\omega} - 
   \sqrt{\left(\frac{M}{1 \pm q\omega}\right)^2 - a(a \mp q)} .
 \end{equation}
 When $q=0$, $r_\mathrm{ex} \ge r_\mathrm{h}$ holds only if $M=a$.
 In this case the equation (\ref{eq:quartic_equation}) becomes
 degenerate and then we have
 $r_\mathrm{ex} = r_\mathrm{eff}^\pm = r_\mathrm{h} = M$.
 Since we are interested in the roots $r=r_\mathrm{ex}$ such that
 $r_\mathrm{ex} \neq r_\mathrm{eff}^\pm$,
 we assume $q \neq 0$ hereafter.
 To satisfy $r_\mathrm{ex} > 0$ leads to $a \mp q >0$ and
 $M/(1 \pm q\omega) > 0$.
 Therefore, we have the following inequality
 \begin{equation}
  a \mp \frac{q}{2} \ge \sqrt{a(a \mp q)} \ge r_\mathrm{ex} , 
 \end{equation}
 where we have used $(A+B)/2 \ge \sqrt{AB}$ for $A=a$ and $B=a \mp q$
 in the former inequality, and $A+B \ge \sqrt{A^2 + B^2}$ for
 $A = \sqrt{a(a\mp q)}$ and $B=M/(1 \pm q\omega) - r_\mathrm{ex}$
 in the latter inequality.
 In addition, $M \ge a$ yields
 \begin{equation}
  M + \sqrt{M^2 - a^2 + q^2} \ge a + |q| > a \mp \frac{q}{2} .
 \end{equation}
 As a result, we have $M + \sqrt{M^2 - a^2 + q^2} > r_\mathrm{ex}$.
 Since $\Delta(r)$ is a monotonically increasing function of $r$ for
 $r>r_\mathrm{h}$,
 we have $q^2 > \Delta(r_\mathrm{ex})$.
 This cannot satisfy the regularity condition (\ref{eq:condition2_app})
 clearly, so that
 we can rule out $r=r_\mathrm{ex}$.

  \subsection{Critical point}
  \label{app:critical}
  
  The rigid rotations in a black hole spacetime generally yield two
  light surfaces, which are the inner light sphere and the outer light
  cylinder.
  However, if the angular velocity becomes 
  sufficiently large, 
  the two
  light surfaces 
  will merge into a single connected light surface. 
  We consider the critical case in which the two light surfaces touch
  each other.

  Now, we focus on the equatorial plane, $\theta=\pi/2$.
  For a given $\omega$, 
  the radii of the light surfaces are
  determined by $f(r,\omega) = 0$, where we have defined 
  \begin{equation}
   \begin{aligned}
    f(r,\omega) \equiv& r F(r,\pi/2) \\
    =& - \omega^2 r^3 + (1-a^2\omega^2)r - 2M(1-a\omega)^2 .
   \end{aligned}
  \end{equation}
  The critical radius $r_\mathrm{c}$
  and angular velocity $\omega_\mathrm{c}$ such that the two light
  surfaces merge are given by the following conditions 
  \begin{equation}
   f(r_\mathrm{c},\omega_\mathrm{c}) = 0, \quad 
   \frac{\partial f}{\partial r}(r_\mathrm{c},\omega_\mathrm{c}) = 0, 
  \end{equation}
  which means that two positive roots of $f(r,\omega)=0$ should be degenerate.

  On the equatorial plane, the radius of the effective horizon,
  $r_\mathrm{eff}$, should satisfy 
  \begin{equation}
   f(r_\mathrm{eff},\omega) = 0, \quad \Delta(r_\mathrm{eff}) = q^2 .
    \label{eq:horizon_condition_equatorial}
  \end{equation}
  Since $f(r,\omega)$ is a quadratic function in terms of
  $\omega$,
  the equation $f(r,\omega) = 0$ has two roots
  \begin{equation}
   \omega^\pm (r) \equiv \frac{2Ma}{r^3+a^2r+2Ma^2}
    \pm \frac{r \Delta^{1/2}}{r^3 + a^2 r + 2Ma^2} .
  \end{equation}
  Note that we find $\omega^+(r)>0$ and
  $\omega^-(r) \le \Omega_\mathrm{h}$ for $r\ge r_\mathrm{h}$. 
  Then, the latter condition $\Delta(r_\mathrm{eff}) = q^2$ yields
  $r=r_\mathrm{eff}(q) \equiv M + \sqrt{M^2-(a^2-q^2)}$, so that 
  we obtain the relation between $\omega$ and $q$ on the equatorial plane
  ($\theta_\mathrm{eff} = \pi/2$) 
  \begin{equation}
   \omega_{\pi/2}(q) =
    \frac{a-q}{2M^2 - q(a-q) + 2M\sqrt{M^2 - (a^2 - q^2)}} ,
  \end{equation}
  as shown in Eq.~(\ref{eq:omega_equatorial}).
  Here, we have taken $\omega_{\pi/2}(q) = \omega^-(r_\mathrm{eff})$
  for $q \ge 0$ and
  $\omega_{\pi/2}(q) = \omega^+(r_\mathrm{eff})$ for $q<0$ in order to
  belong to the $r^+_\mathrm{eff}$ branch. 
  Differentiating Eq.~(\ref{eq:horizon_condition_equatorial})
  with respect to $q$ and evaluating them  
  at the
  effective horizon, we have the following identities   
  \begin{equation}
   \partial_r f \frac{dr_\mathrm{eff}}{dq}
    + \partial_\omega f \frac{d\omega_{\pi/2}}{dq} = 0 ,\quad
    \Delta_{,r} \frac{dr_\mathrm{eff}}{dq} = 2q .
  \end{equation}
  As a result, we have 
  \begin{equation}
   \frac{d\omega_{\pi/2}(q)}{dq}
    = - \left.\frac{2q \partial_r f}
	 {\Delta_{,r} \partial_\omega f}
	\right|_{(r,\omega)=(r_\mathrm{eff},\omega_{\pi/2})} .
  \end{equation}
  Because $\partial_r f=0$ at the critical point $r=r_\mathrm{c}$ as we have seen, we 
  have proven that if the effective horizon coincides with the
  critical radius, namely $r_\mathrm{eff} = r_\mathrm{c}$, then
  $d\omega_{\pi/2}(q)/dq = 0$.

  \subsection{Monotonicity and positivity}
  \label{app:positivity}
  
  Now, we shall prove $1-\omega a \sin^2\theta > 0$ at the effective
  horizon $r=r_\mathrm{eff}^\pm$.
  We have 
  \begin{equation}
   \frac{\partial}{\partial\omega}[\Delta(r^\pm_\mathrm{eff}) - a\omega q^2]
    = \mp q
    \left\{
     \Delta(r^\pm_\mathrm{eff}) +
     \frac{2M r^\pm_\mathrm{eff}}{(1 \pm q\omega)^2}
     \left(r^\pm_\mathrm{eff} - M
      \frac{1 \pm q\omega}{1 \pm q\omega+q^2\omega^2}\right)
     \left(r^\pm_\mathrm{eff} - \frac{M}{1 \pm q\omega}\right)^{-1}
     \left[\left(\frac{1}{2} \pm q\omega\right)^2 + \frac{3}{4}\right]
    \right\},
    \label{eq:monotonicity}
  \end{equation} 
  where each term in 
  the braces 
  is positive.%
  \footnote{
  $r_\mathrm{eff}^\pm \ge M \ge
  M\left(1 \pm q\omega\right)/\left(1 \pm q\omega+q^2\omega^2\right)$ 
  }
  Thus, $\Delta(1-\omega a\sin^2\theta)$ at $r=r^\pm_\mathrm{eff}$ is a
  monotonic function of $\omega$ for a fixed $q$, and whether it is
  monotonically increasing or decreasing depends on the sign of $q$.
  Since $M \ge a$, we have $a^2 - q^2 - 2M^2 < 0$.
  It leads to
  \begin{equation}
   \begin{aligned}
    & a^2 - q^2 - 2M^2 < 0 < 2M\sqrt{M^2 - a^2 + q^2} \\
    \iff& a(a-q) < 2M^2 -q(a-q) +2M\sqrt{M^2 - a^2 + q^2} .
   \end{aligned}
  \end{equation}
  Because the right-hand side of the last inequality is positive if $q<0$,
  we obtain 
  \begin{equation}
   \omega_{\pi/2}(q) =
    \frac{a-q}{2M^2 - q(a-q) + 2M\sqrt{M^2 - (a^2 - q^2)}} < \frac{1}{a}
    \quad \text{for}\quad  q<0.
  \end{equation}
  Furthermore, as we have shown in the previous subsection, we obtain
  \begin{equation}
   \omega_{\pi/2}(q) = \omega^-(r_\mathrm{eff}) \le \Omega_\mathrm{h}
    \le \frac{1}{2a} < \frac{1}{a} \quad
    \text{for} \quad q \ge 0 .
  \end{equation}
  As a result, we conclude that
  \begin{equation}
   [\Delta(r^\pm_\mathrm{eff}) - a\omega q^2]|_{\omega=\omega_{\pi/2}(q)}
    = q^2 [1-a\omega_{\pi/2}(q)] > 0 .
    \label{eq:positivity}
  \end{equation}
  Note that, even though $\omega_{\pi/2}(q)$ has been defined only for the
  $r^+_\mathrm{eff}$ branch, the above result can be applied to the
  $r^-_\mathrm{eff}$ branch because the $r^\pm_\mathrm{eff}$ branch is
  related to each other $q \to -q$. 

  From Eqs.~(\ref{eq:monotonicity}) and (\ref{eq:positivity}),
  we have proven that $1-\omega a \sin^2\theta > 0$ at
  $r=r_\mathrm{eff}^\pm$ for arbitrary $q$.

\end{document}